# Advanced Traffic Management Systems: An Overview and A Development Strategy

M. Shahgholian[1], D. Gharavian[1*]

1: Department of Electrical and Computer Engineering, Shahid Beheshti University, Tehran, Iran
*D_Gharavian@sbu.ac.ir

**Abstract:** Nowadays the problem of traffic congestion is known as the main cause of air pollution in the cities of the world. Urban traffic engineers and managers have proposed three general approaches to dealing with this phenomenon. The first approach expands the capacity of the urban traffic network (UTN). This approach cannot be implemented in many urban areas due to urban density and traffic volume. Further, implementation of this approach, in addition to its high cost, requires an accurate understanding of the dynamic and static properties of UTN for well planning. The second approach can be called the traffic assignment. In this approach, through software applications or information boards, network managers are informed about the users' status in the network and offer the best route to them. Finally, the third approach involves optimizing the capacity of the UTN. This approach tries to control the traffic actuators in order to create the maximum capacity for network users. Maximum capacity means that the total amount of urban trip times is minimized or the total flow of the network links is maximized. Therefore, it can be said that the advanced traffic management systems (ATMS) are based on four main sections. These four sections include traffic information, traffic assignment, traffic optimization, and traffic prediction. This paper initially presents an overview of these four sections, in the end, it proposes a development strategy for the ATMSs.

## 1-1. Introduction

The main difference between the phenomenon of traffic and other social phenomena is its reverse growing. Simply, with the improvement of social indicators such as security, well-being, and economy, urban trips increase thereby developing traffic congestion. With the increasing urbanization and cheaper access to vehicles, traffic congestion is a major recurring problem in many large cities. Traffic congestions cost us time, money and health. Also, vehicles in congestion condition can burn up to 80% more fuel than those in free traffic, and this leads to more air pollution[1]. According to data released in 2010, cars annually emit about 5.53 million Gg carbon dioxide into the atmosphere, which is 16 percent of the world's total[2]. Motor vehicles also produce 72 percent of nitrogen oxide and 52 percent of reactive hydrocarbons in the world[3]. These greenhouse gases by trapping heat in the atmosphere, causes worldwide temperatures to rise that was 0.71 C in 2017. Warmer global atmosphere affects farming, wildlife, sea levels and natural landscapes. Also, the health risks of air pollution are extremely serious. Poor air quality causes respiratory diseases, such as asthma and bronchitis, and the risk of developing dangerous illnesses such as cancer, are increased, and the health care system faces substantial medicinal costs. Suspended particles alone each year are responsible for about 30,000 premature deaths in the world[4].

On the other hand, the report by transport analytics company Inrix found that drivers in Los Angeles spent an average of 102 hours waiting in traffic last year. Moscow followed with 91 hours lost to congestion, while London and Paris lost 74 and 69 hours respectively[5]. On average, congestion costs in LA is $19.2 billion for the city as a whole. Costs are higher in New York at $33.7 billion. This cost for total London vehicle commuters is $12.2 billion and for Berlin is $7.5 billion. The economic damage of congestion last year in the United States, Germany, and Britain totaled US$461 billion. Such costs are rising as the world's population and urbanizes grows[6].

These health threats and economic costs of UTN congestions have led engineers to design different solutions to deal with them. One of the options for reducing congestion is developing transportation infrastructure in proportion to the number of increased vehicles. This method is largely infeasible in developing countries where space is a major constraint. The difficulty of increasing the capacity of UTN, especially in urban areas, and the continuous growth of urban trips in the time unit, have caused the traffic congestion phenomenon to be the main problem of network managers. Inability to adding capacity in line with the demand growth has led engineers to try other solutions. All of these solutions have two main aims that are decreasing the travel time of UTN commuters and use the network capacities with the maximum efficiency[7]. These aims are achieved by traffic assignment and traffic optimization. In other words, all solutions can be categorized into these two sections. The process of allocating a given set of trip interchanges to a specified transportation system is usually referred to as a traffic assignment. On the other hand, these sections have a hierarchical effect on each other[8]. This effect necessitates the coordination between these two parts. In other words, an accurate development strategy to implement these sections should be designed. This process concludes a system that called ATMS.

Through the deployment of sensing, communications, and data-processing technologies, implementation of the ATMS is expected to be more facilitated. Indeed, ATMS seek to reduce traffic congestion in UTN by improving the efficiency of utilization of existing infrastructures.

This paper is structured as follows: Section two presents different methods of obtaining traffic information in manual and automatic sections. Section three deals with the essential parts of traffic assignment, presented in three subsections, covering network loading model, trip matrix, and travel choice principle. Based on that, the organization of traffic assignment implementation is presented. Section four discusses traffic network optimization. It divides UTN into road and freeway networks. Then, we try to propose how to optimize these networks with traffic actuators such as traffic signal and ramp metering. The steps of applying traffic optimization are expressed at the end of this part. Section five proposes different methods of traffic prediction for long and short forecasting. Long forecasting is used in traffic assignment, while traffic optimization employs short forecasting. Prediction methods are categorized into naive, parametric and nonparametric methods. The final section combines these four sections and concludes a development strategy to implement the ATMS.

## 1-2. Traffic Information

In recent years, technological developments have enabled collection and transition of real-time traffic information, and this, coupled with the increase of traffic trips and congestion, has increased the interest in traffic modeling [9]. These information and models are the main bases of ATMSs. Currently, a wide range of methods has been proposed to obtain traffic information. According to studies, these methods can be divided into manual and automatic approaches.

### 1-2-1. Manual

The most common methods for collecting traffic data are the manual methods. These methods can be classified into manual counting and survey.

#### 1-2-1-1. Manual counting

A manual count is performed by one person using a counting board to count all vehicles at the intersection or selected paths for a predetermined period of time. This method is also used to collect data in order to determine the classification of vehicles, the percentage of turning at intersections, pedestrian movement, or vehicle occupancy.

Manual counts use one of these three methods to record data [10]: tally sheets, mechanical counting boards, and electronic counting boards.

**Tally Sheets:** The data can be recorded with a check sign in a pre-prepared form. A chronometer or stopwatch is required to measure the desired time interval.

**Mechanical Counting Boards:** Mechanical counting boards include counters mounted on a page which record each line of the road. The common forms of these boards include pedestrian, bicycle, vehicle classification, and traffic volume counts. Typical counters are push-buttons with three to five registers. Each button is used to count a specific type of vehicle or pedestrian. A chronometer or stopwatch is also required to measure the desired time interval.

**Electronic Counting Boards:** Electronic counting boards are tools used to collect traffic counting data which use the battery. Electronic counts are more compact and easier to use than mechanical boards. They are supplied with battery and have an internal clock that automatically separates the time intervals of data collection. Also, the data can be uploaded to a computer, which saves time and decreases the human errors.

This method of data collection can be expensive in terms of manpower. However, in cases where it is necessary to collect data related to different vehicles classification separately, or when the infrastructures of automatic methods are not implemented, these methods should be used [11].

### 1-2-1-2. Survey

Traditionally, UTN managements use household questionnaires or road surveys to estimate the origin-destination matrix (ODM)[12]. The correct estimation of the ODM with respect to the collected data requires experience, skill, and proper understanding of the study area. It is also important to know the purpose of the study and the detail of modeling methods since the data are affected by these properties. Further, many practical considerations including the availability of time and money have a great impact on the survey design.

- **Different Types of Surveys**

Household surveys are considered as one of the most reliable methods for obtaining information about urban travel patterns. In this method, ideally, the trip information of all individuals in the study area must be obtained to determine the pattern of trips. This is not possible due to the need for a large number of resources and time. Moreover, this will cause problems in handling large data at the modeling stage. Therefore, samples from households are randomly selected and a survey is conducted in order to obtain the required data. The sample size is determined based on the population size. Typically, for populations under 50,000, at least 10% of houses are required for sampling. But, for populations of more than one million, only 1% is needed to achieve the same precision [13].

In addition to household surveys, other studies are needed to calibrate and validate models or actions as complementary. These include origin-destination (O-D) surveys[14], roadside interviews [15], and cordon and screen line counts [16].

While travel surveys provide extremely useful data in order to formalize and estimate behavioral choice models (e.g. the choice of a destination or mode of transportation), they are much less useful for constructing ODM due to an inadequate number of trips in many matrix elements. Furthermore, surveys have increasingly been dealing with some issues at the stage of creating sample data. Issues such as falling response rates and unreported trips reduce the quality of estimated ODM. Therefore, other types of resources are also used to generate the trip matrix, such as roadside traffic counts, cordon or screen-line surveys, and public transport surveys. The resulting data are then used for improving the quality of ODM, though they do not always contain the necessary information. This applies, for example, to road traffic counting, which provides traffic volume information at a specific point on a road, not the origin and destination of the trip [17]. So, this approach has two main drawbacks [12]:

- Calculating an ODM from the initial data collection to exploitation of the first results can take years and produce only a snapshot of the travel demand;
- The collected data are limited in both spatial and temporal terms.

### 1-2-2. Automatic

Automatic methods gather data from detectors that sense network conditions. Traffic sensors are divided into point and interval detectors.

#### 1-2-2-1. Point detectors

These types of detectors are located at fixed points of the street and record traffic information at these specific points. These sensors accurately capture the data and are not influenced by external factors. They have widely been deployed on the streets, but their installation and maintenance are costly and complex. For this reason, in recent years, other solutions such as video image detection techniques have been used thanks to their low installation costs and high precision. In these methods, video cameras and image processing are used to obtain the number of vehicles and speeds at specific points of the road. Their main drawback is that they are usually sensitive to external factors, such as weather, and also need periodic maintenance [18].

#### 1-2-2-2. Interval detectors

Interval detectors capture data that enables the direct calculation of travel time between two points, as opposed to point detectors that are only able to describe a single point of the road. This type of detector can be further divided into two main groups: floating or probe vehicles and automatic vehicle identification (AVI) techniques [18].

### ❖ Floating Car Data (FCD) or Probe Vehicles

Floating and probe vehicles are a sample of vehicles circulating in the traffic network and providing information about their trajectories. The main difference between them is that floating vehicles are specifically employed for data collection purposes, while probe vehicles are passive vehicles that travel in the UTN for other reasons. Both of them are generally equipped with cell phones and/or Global Positioning Systems (GPS), and every few seconds, they send the position, direction, and speed information to the UTN management center. In this method, the vehicle path and travel times between its O-D can be achieved easily and reliably [18]. On the other hand, the wide deployment of pervasive computing devices (mobile phone, smart cards, GPS devices, digital cameras, and so on) has provided unprecedented digital footprints, telling where people are and when they are there [12]. In other words, the FCD principle involves collecting real-time traffic information by tracking vehicles based on data received from cellular phones or GPS embedded in UTN vehicles. This basically suggests that any vehicle equipped with a mobile phone or GPS acts as a sensor in the UTN. After collecting and processing data, useful information (such as traffic conditions and alternative routes) can be provided to managers and users of the UTN [19].

Based on the type of sensor, FCD can be divided into GPS-based and cellular-based systems[19].

#### 1- GPS-based

Since GPS data are able to provide useful data with a specific precision, in the mid-1990s, researchers began to explore the possibility of obtaining trip data from GPS data. These GPS devices still have disturbances such as lack of response to the signals of devices or losing access to the GPS signal in the building, underground, tunnel and urban canyons areas [20].

Since GPS function is attached to the smartphones, some researchers have also started to use smartphones' GPS data to identify personal trip information. They also combine this with a web-based diary system or a Geographic Information System (GIS) to receive additional and confirmation information of transportation modes and trip purposes[21]. Additionally, the assisted GPS system, called AGPS, is widely available on smartphones. This technology can receive GPS signals inside buildings, vehicles, urban canyons (where tall buildings and other edifices block GPS signals). This system has improved and enjoys a highly sensitive receiver. This means that more accurate GPS data can be obtained by smartphones[22]. If the technologies of automatically deriving personal trip data can also be achieved with higher accurate results, GPS data collection through smartphone may become the main method of personal trip data collection in the future with lower costs and minimum burden on respondents[20].

#### 2- Cellular-based (e.g. CDMA, GSM, UMTS and GPRS networks)

Mobile phone operators, who, for legal or billing purposes, are obliged to record information about the use of these devices, find themselves with increasingly informative databases. In other words, each time a mobile terminal is used for a call or for sending an SMS (Short Message Service), the operator records the call features, including the timestamp, the base station's identifier too which the user is connected, and quantitative data (call duration, volume of data exchanged) [17].

Unlike fixed traffic detectors and GPS-based systems, in cellular-based approach no special device is required for the vehicles and no specific infrastructure is built around the roads. Therefore, it is cheaper than conventional detectors and provides greater coverage potentials. Traffic data are obtained continuously instead of isolated point data. It is faster to set up, easier to install and needs less maintenance [19].

Meanwhile, the mobile phone trace data have also significant drawbacks for transportation research [23]:

- Socio-economic and demographic characteristics are not available due to privacy concerns. These features are necessary for the calibration of models at a coherent level, in order to discover mechanisms of behavior based on individual/household trips;
- Mobile phone users may not represent a proper sample of the population. The results require careful and correct analysis interpretation;
- These datasets are not originally designed for modeling purposes and are often not easy to use, limiting the usefulness of raw data without preliminary processing;
- The localization data associated with each log is limited to the position of the base station used, which results in a positioning uncertainty ranging from approximately a hundred meters in a dense urban zone to several kilometers in rural zones.

Today, ride-sharing services such as Uber[24] have been attracting a great deal of attention. Compared to traditional taxi services, users travel more easily, faster, and cheaply in the UTN. Smartphones are a key tool for these ride-on-demand services. Firstly, passengers use mobile phones to locate both themselves and drivers in their request for rides. Their locations, behavior (e.g., estimating trip fares, finding nearby drivers, creating orders, using particular features, etc.) and relevant information are recorded by the service provider. Secondly, drivers have their locations uploaded periodically by the on-car mobile GPS devices. In short, the use of mobile devices (including mobile phones and GPS devices) enables the tracking of locations and behavior of drivers & passengers. It also opens an in-depth study field of such a service from different perspectives [25].

- **Automatic Vehicle Identification (AVI)**

AVI refers to the technology used to identify a particular vehicle when it passes a particular point. Early development of AVI occurred in the United States, beginning with an optical scanning system in the 1960s to identify railroad boxcars [26]. AVI systems can be of various types, such as automatic toll collection systems [27], vehicle-mounted transponders of different types and roadside beacons [28], video cameras and license plate matching techniques [29], and the more recent Bluetooth and WIFI based detection systems [30]–[32]. These devices detect and identify vehicles only at the beginning and at the end of the study segment and calculate the travel time directly from these data.

AVI can serve a range of purposes to: charge for road use, suggest routes for drivers, improve UTN management, detect stolen vehicles, and monitor fleets of trucks, buses, and taxis [26]. Developments in AVI and monitoring technology have opened the possibility for a system of road pricing in which charges depend on the time of day, road, and vehicle. This could include possible higher charges for travel, for example, during rush hours and on specific congested roads. According to neoclassical economics, this would improve the efficiency with which the roads are used [26].

Interval detectors guarantee high accuracy and quality description of the traffic situation. However, there are some practical inconveniences when using them. For example, many interval detectors are unable to identify all vehicles on the UTN, and therefore an accurate statement of the identified vehicles is an issue that should be considered. The sample size needed for accurate representation of traffic is generally quite large and difficult to obtain in real situations, especially in the case of probe vehicles [18]. In addition, some interval detectors, such as GPS-equipped vehicles, provide irregular intermittent data. Modeling this type of data is more complicated due to the uncertainty and lack of the information associated with the irregularity of the sampling. Because of this, most travel time models are aimed at regularly spaced data [20].

**Fig. 1** demonstrates the classification of traffic information.

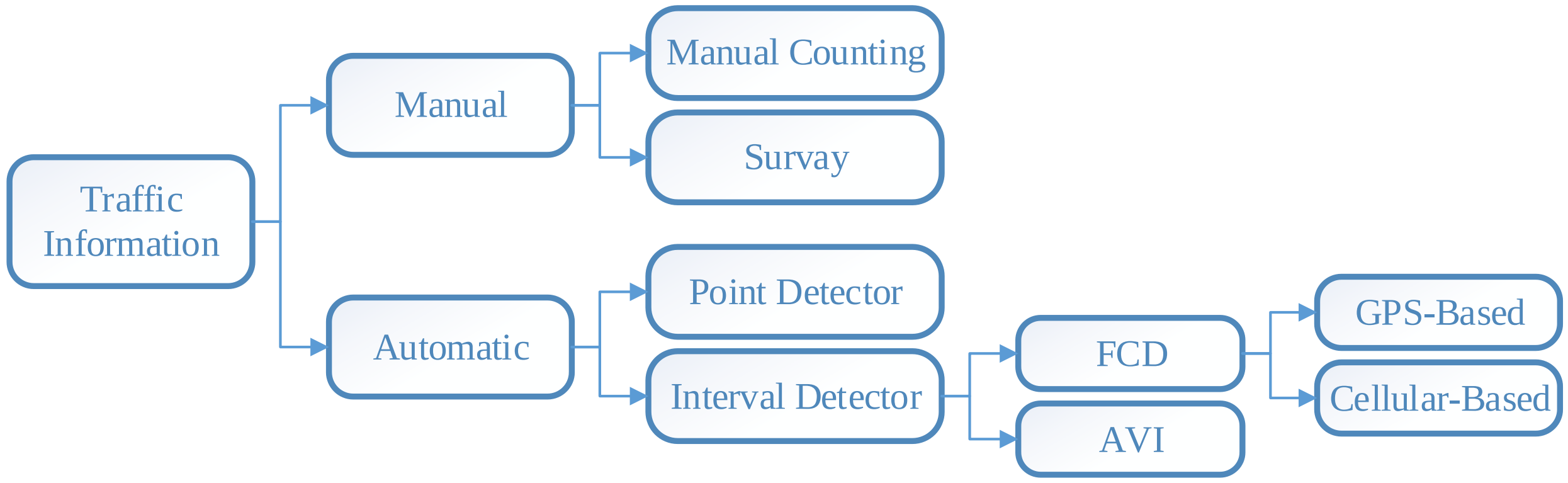

**Fig. 1:** Methods of obtaining traffic information

## 1-3. Traffic assignment

The traffic assignment problem seeks to determine flows on links of a given road network under certain optimality or equilibrium conditions, where a flow rate is given for OD-pairs of nodes for every point considered in time. The departure flow rate for an OD-pair at time t determines the amount of flow leaving the origin node to travel to the destination node at time t. When the given flow rate for the OD-pairs is constant over a long time, the traffic assignment problem is called static, otherwise, it is dynamic[33]. In other words, these models determine the flow on each link and capture the interaction between demand and supply. The traffic assignment models are crucial for traffic flow and travel time forecasting in long-term transportation planning, as well as in short-term traffic operation management and control[34].

The major aims of traffic assignment procedures are [35]:

1. Estimating the traffic volume of network links and obtaining the overall network values
2. Estimating the cost of the trip between the zones
3. Determining the trip pattern of each O-D pair
4. Identifying congested links and collecting useful traffic data to design UTN's control plan

With regard to these aims, both static and dynamic traffic assignment can be divided into three main components called the network loading model, the trip matrix estimation, and the travel choice principle.

### 1-3-1. Network loading model

Also known as the traffic flow component, the network loading model explains how traffic is distributed inside the road network, based on which the network performance is determined in terms of travel time[36]. Modeling the traffic flow components as a unique mapping of the route flows creates a new way of analyzing traffic assignment problems with path-time outcomes. This approach has two advantages: first, it ensures the proportionality between link travel time and link flow because link travel time is uniquely derived from the link flow. Second, it allows us to determine a unique response to the traffic assignment problem directly by simply checking whether the unique mapping is continuous or strictly monotonic[37].

The network loading model can be divided into two parts: link model and node model. The outputs are affected by the types of link and node model in traffic assignment. Based on the level of detail in the representation of UTN features, network loading models are typically classified into macroscopic, mesoscopic, and microscopic[36]. The following discusses how the optimal ODM or node model is obtained. The following traffic behavior can be considered in the link model[37]:

1. First-in-first-out (FIFO): FIFO on the link level means that users who enter the link earlier will leave it first;
2. Causality: Causality means that the speed and travel time of a vehicle in a link is only influenced by the speed of the front vehicles;
3. Queue spillback: Queue spillback refers to the end of queue spilling backward in the network.

The above traffic behavior governs the properties of traffic assignment formulations such as the properties of the route and O-D costs and solution properties (e.g., the existence of solutions). A variety of mathematical methods are considered for network loading models such as the dynamic maximum flow problem, the earliest arrival flow problem (also known as universal maximum flow problem), the quickest flow or the transshipment problem, and a dynamic minimum cost flow problem[33].

### 1-3-2. Trip Matrix/Origine-Destination Matrix(ODM)

Any movement from one point (origin) to another point (destination) of the city which is done to perform something (the purpose of the journey) is called "trip". Knowing the magnitude of attracted and generated trips from each zone is essential for planning and managing UTNs. In order to determine the ODM, initially different geographical regions of the network should be analyzed. Each geographical region is defined as a zone and in order to model the graph of traffic network, a node is considered for each zone. The graph of the network is constructed via connecting nodes with links. Links are models of the routes between every two zones. Trips are generated through the displacement of the traffic network users between zones. Choosing the number and size of zones requires a balance between the simplicity and the precision of network modeling. This means that if a low number of zones are considered, the analysis will be simpler. On the other hand, as the number of zones grows, more trips can be extracted.

Traditional methods for estimating ODM used large-scale sampled surveys such as household survey, roadside interview, and license plate method conducted once in every 1-2 decades. Nevertheless, these surveys become impossible to conduct due to financial constraints. Also over time, the survey data become obsolete to obtain ODM[38]. Consequently, since the late 1970s, many models have been proposed and widely applied for estimating/calibrating sampled/old matrices using the current data of traffic counts collected on a set of links. The accuracy of this estimated matrix depends on the input data errors, the estimation model, and the utilized calibration/optimization method[39].

#### 1-3-2-1. Dynamic and Static ODM

The static ODM estimation does not consider the time-dependent traffic flows and is assumed to state steady state during a period of time. The average traffic counts are collected for a longer duration to determine the average O-D trips[38]. The dynamic estimates of time dependencies in ODMs are a major input to dynamic traffic models used in an ATMS for estimating the current traffic state as well as for forecasting its short-term evolution. Travel time forecasting and dynamic ODM estimation are two key components of ATIS and ATMS[32]. The main difficulty of dynamic ODM estimation and prediction arises from the following characteristics:

❖ **Underdeterminedness Problem**

Many combinations of demand patterns can result in the same link-flow values; thus, the problem of estimating the ODM from traffic counts is typically underdetermined. The set of possible solutions usually grows with the size of the UTN, with more routes being available for each O-D pair. However, this usually decreases with the addition of information sources [39].

❖ **Nonconvexity Problem**

In congested networks, the relationship between the link flows and the ODM causes a high nonlinearity over time, mainly because of the spatial and temporal dynamics of queues and delays, spillback and rerouting effects, and corresponding changes in the split proportions at the nodes. This dependency makes the problem to be highly non-convex[39]. Therefore, the existing dynamic O-D estimation models have to be solved as a bi-level program. The procedures available for this have limitations which can be summarized in two parts[40]:

Firstly, because of the nonconvexity mathematical programs of this problem, it is hard to formulate it appropriately and solve it effectively.

Secondly, to model the traffic congestion condition, we may reasonably anticipate the need to represent nonlinear spatiotemporal dynamic processes. This means that the dynamic O-D prediction models would be more useful if the traffic flow module can represent accurately sudden falls in speed and flow with consequent delays which are the main features of congested traffic. For predicting the congested traffic, certain reproducible patterns are observed, including where bottlenecks are located, when they become activated and what their special features are. Spatio-temporal traffic features do not change significantly from day to day, such as where the onset of queuing was marked by a sudden fall in flow and increase in measured occupancy. Also, the speed of the spillback shock determining the onset of queuing is similar among days.

### 1-3-2-2. ODM Estimation Methods

The methods proposed for modeling the ODM can be broadly assigned into three groups:

1. **Gravity Model**

Initially, the researchers tried to estimate the ODM as a function (like the gravity models) with related parameters. Most of the proposed methods and all the applications with real data fall into gravity models. The modeling under this category can be presented with several degrees of sophistication, leading us to consider two sub-groups [41]:

- Gravity models leading to linear equations on the links
- Gravity models leading to non-linear equations on the links

Some of the researchers used the Gravity model and some used Gravity-Opportunity (GO) based model for ODM estimation. These techniques need trip data between every two zones for calibrating the parameters of the demand models. The GO models were found to consume more time than the GR model and did not guarantee the reliability of the estimated matrix[42].

2. **Entropy Maximizing (EM)/Information Minimizing(IM) Approach**

The concept of entropy originated in physics. In a closed physical system, its elements tend to form an arrangement which can be organized in as many ways as possible compatible with the system constraints (energy, mass, etc.)[43]. This arrangement is the most likely one with the greatest "disorder". The idea of entropy is also linked to that of "information". It can be seen intuitively that at least the state of the maximum disorder is the one requiring a minimum volume of information. The potential information content of a message grows as the sequence of symbols departs from a purely random (high disorder) sequence.

The second group of models could be referred to as EM or IM approaches. These techniques are used as model building tools in transportation, urban and regional planning context after Wilson [44] introduced the concept of entropy in modeling these systems (systems with large numbers of components with apparent disorganized complexity). This approach can be used if we want to find the most likely ODM compatible with the available set of link counts, i.e., if we intend to "exploit" all the information contained in the observed link flows to determine the most likely ODM compatible with them [41].

Summary the EM/IM procedure analyzes the available traffic information to obtain a unique probability distribution. This method is useful for the under-determinedness problem proposed in the previous section. However, unfortunately, IM- and EM-based models have the disadvantage of not considering the uncertainties in traffic data and primary matrix, which can be incorrect and change the output [38]. Clearly, this is not desirable for a nonconvex problem.

### 3. Equilibrium Approach

This approach attempts to estimate ODM through a network equilibrium method based on Wardrop's first principle. The ODM estimation methods developed for the networks with no congestion basically assume that the route choice step is independent of the ODM estimation process. This assumption is very useful for simplifying the mathematical analysis, but it is somehow far from the real UTN. Therefore, some researchers have considered congestion effects in the ODM estimation problem. In other words, the dependence of the link costs, path choices and assignment fractions on link flows should not be neglected. Equilibrium assignment approaches are particularly adaptable for this nonconvex problem [45].

Sang Nguyen has extended some of Robillard's ideas for the equilibrium assignment case, which is more relevant to urban areas. Nguyen suggests two methods. The first belongs to the cases where traffic information is available for all links. The second method requires only costs for all O-D pairs to be known and does not explicitly ask for traffic information at all links. The first case is only suitable when obtaining traffic information for all links is possible. This would restrict its application to small networks. In addition, the algorithm is likely to be slow for a large number of O-D pairs. Nguyen then proposed the second method which reduces the data requirements. It is interesting to note that Nguyen has only identified the necessary conditions to solve the problem and not the sufficient ones. This suggests that the solutions found with his method meet the equilibrium conditions but they will not be normally unique solutions [41]. In other words, his method is not appropriate for underdetermines problem.

## 1-3-2-3. Optimization Methods for ODM Estimation

Several models have been presented to calibrate or update ODM from traffic counts for networks via parametric estimation techniques. The most usable forms of these methods are:

### 1. Bayesian Inference (BI)

The BI approach considers the target ODM as a prior probability function of the estimated ODM. If the observed traffic counts are considered as another source of traffic information, the Bayes theorem provides a method for combining two sources of information [46].

### 2. Generalized Least Squares (GLS)

Among the branches of regression analysis, the GLS estimation method, based on known Gauss-Markov theory, plays a fundamental role in many theoretical and practical aspects of the statistical inference-based model. The advantage of this method is that no distributional approximations are assumed in the data sets. This allows combining survey data (which is directly related to O-D motions) with traffic count data, while the relative accuracy of these data is also considered [45].

### 3. Maximum Likelihood (ML)

The ML approach is one of the oldest and most important methods in estimation theory. The ML tries to compute the set of parameters that generate the most frequent observed sample. In the ODM estimation, the ML method maximizes the probability of observing the target ODM according to the observed traffic counts data to achieve the true trip matrix [47], [48].

### 4. Kalman Filter

The Kalman filter algorithm[49] has been widely proposed to cater for real-time requirements. This algorithm solves a least-square problem in an incremental fashion, allowing the solution to be updated when additional data are available [50].

The estimation process using Kalman filter can be divided into two steps: prediction step and correction step. The prediction equations in the Kalman filter use the current state of the system and covariance of this time-step based on the past state to obtain a prior estimate. The correction equations provide a feedback on the prediction by incorporating a new measurement to update a priori estimate and obtain an improved posterior estimate of the process state. Kalman Filter is an optimal state estimation process applied to a dynamic system which offers a minimum error variance and linear unbiased recursive algorithm. The problem of estimating ODM is formulated using the linear map of a state variable to a measurement variable. The applications of Kalman filter are limited as it does not estimate the state of a nonlinear and non-Gaussian system [51].

The extended Kalman filter (EKF) was proposed to estimate the state of nonlinear dynamic systems. The EKF linearizes the estimation of the current mean and covariance by taking partial derivatives of the process and measurement functions into account using the Jacobian matrix. The process state in the extended Kalman filter is approximated by Gaussian random variable and it has been considered analytically through the linearization, which could corrupt the mean and covariance of the state estimate. Unscented Kalman filter (UKF) provides a derivative-free approach to estimating the state of a nonlinear process. UKF selects a set of points deterministically to estimate the mean and covariance of the nonlinear dynamic process [51].

### 5. Particle Filter

Gordon et al. (1993) proposed particle filter (PF) based on sequential Monte Carlo method, which provides an effective solution for nonlinear and non-Gaussian problems. Particle filter also has an advantage over UKF, as the number of sigma points in UKF is based on an algorithm and is far smaller than the number of sample points selected in the particle filter. The estimation error in the UKF filter does not converge to zero, however, in the particle filter algorithm, the error can converge to zero with an increase in the number of sample particles.

Nevertheless, it needs analysis of a large number of particles, where the optimal numbers of particles cannot be determined. This deficiency makes particle filter infeasible for real-time applications of complex dynamic systems [51].

### 6. Genetic Algorithms

Genetic algorithms derive a globally optimal solution from iterative processes of testing and modifying of a set of solutions. These maintain several good solutions and generate new ones according to designated strategies of combining existing solutions and occasionally introducing arbitrary modifications. They have been found to lead to global solutions by searching over a wide scope [40].

Different ODM approaches are shown in **Fig. 2**.

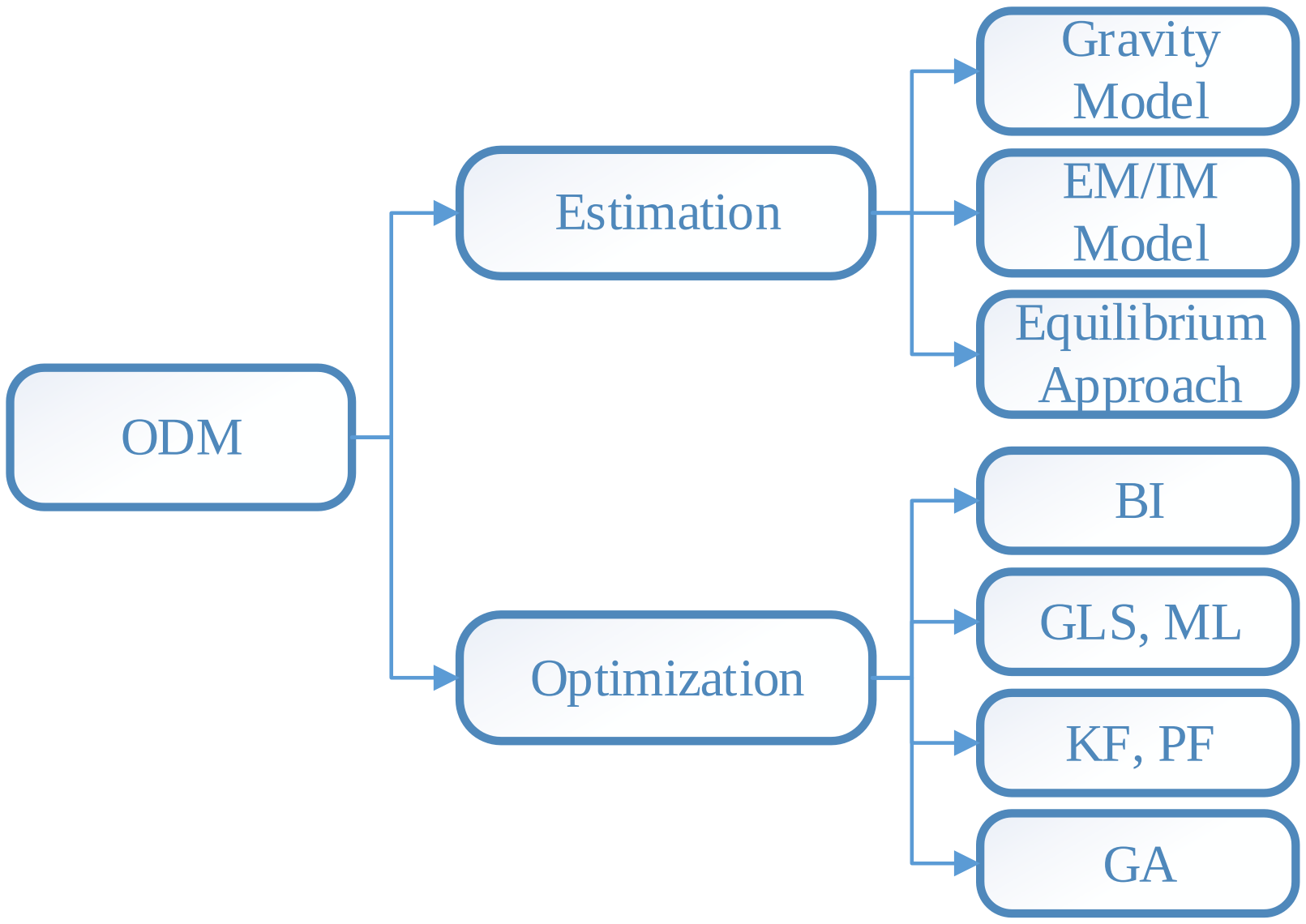

**Fig. 2:** ODM determination methods

## 1-3-2-4. Organization of a dynamic ODM prediction model

The proposed dynamic ODM prediction model consists of the dynamic traffic prediction step, dynamic traffic flow step, and dynamic ODM estimation step. First, the dynamic traffic prediction model estimates the traffic condition and relations of traffic variables (flow, speed, density) which would be an outcome of an upon certain ODM. Then, the dynamic traffic model calculates a link distribution proportion at each time interval and proceeds to input data of the dynamic O-D estimation model. Finally, the dynamic ODM estimation model estimates link flows based on a link distribution proportion at each time interval and estimates dynamic ODM demands to minimize the error between the observed and estimated link traffic flows [40].

### 1-3-3. Travel choice principle

A travel choice principle indicates how the travelers choose their routes, departure times, modes and destinations, i.e. it models the propensity of travelers to travel [37]. It is not easy to predict the user's route choice. The factors that make the answer difficult are:

1. The effect of each choice on the next choices
2. The incidence of unexpected events such as traffic accidents
3. Random behavior of each network user to route choice

With regard to these nondeterministic parameters, travel choice models are separated into two groups: stochastic and nonstochastic methods.

In the Wardrop's first principle, it is assumed that each user has the correct understanding of the shortest path. This means that, in addition to knowing the length of paths, users should know the decision of other network users about their route. In this model, assumed users cannot change their decision. The correct route choice for all drivers results in the equilibrium traffic flow pattern in the entire UTN[52]. If for each O-D pair and at each instance time, the actual travel time for users who are reaching to their destination at that time is equal or less than that, then it can be said that the dynamic traffic flows onto the UTN have reached user equilibrium or dynamic user optimal.

The Wardrop's first principle cannot model the UTN conditions accurately. This model omits random components in modeling the route choice of travelers. Additionally, the extent of demand is not deterministic. So because of these two reasons, researchers have used the stochastic approach to travel choice modeling. With regard to stochastic methods, routes that have minimum stochastic travel times have a greater chance to be chosen. Almost all of these models assume that drivers' perception of costs on each given route is not accurate and trips between each O-D pair are divided between the routes with the cheapest cost. Proposing an analytical model for these approaches depends on the type of random components model. The most general random models are the Normal and Gumbel distribution that is used in Probit and Logit travel choice models, respectively. Other distributions like Gamma and Log-Normal have also been used in articles [53].

Further, travel choice models can be separated into descriptive and normative models. The descriptive formulation attempts to capture how the users behave in the face of a set of traffic conditions (user optimal). Normative models seek to determine how the system should behave in order to optimize some system-wide criteria (system optimal) [54]. For example, static user equilibrium (UE) problem is a descriptive method, while system optimal (SO) problem is a normative method. The SO assignment is based on Wardrop's second principle, which states that drivers cooperate together to minimize the total travel time of the UTN. This assignment can be considered as a model which minimizes congestion when drivers are told which route to use. This method tries to optimize the traffic by minimizing the travel costs, thus achieving an optimum social equilibrium [55]. The other types of static route choice principle include all-or-nothing assignment (AON), incremental assignment, capacity restraint assignment, stochastic user equilibrium assignment (SUE), etc. [56].

Typically, the travel choice principle of dynamic traffic assignment (DTA) is the dynamic extension of the route choice principle of static traffic assignment (STA). The first mathematical programming approach to modeling this problem is the Dynamic System Optimal (DSO) method derived from Wardrop’s second principle [57]. Their model was formulated as a discrete-time, non-linear, non-convex mathematical program, and the corresponding algorithm solved a piecewise linear version of it. Congestion was treated explicitly using conventional link performance functions. Since the formulation was nonconvex, global minimization was achieved through a one-pass Simplex algorithm[58] based on the nonconvex assumption on the cost function. Other methods include the Dynamic User-Optimal (DUO) principle extended from Wardrop’s first principle[59], the Dynamic Stochastic User Optimal (DSUO) principle extended from the stochastic extension of Wardrop’s (1952) first principle[60], etc. These principles assume that travelers select their routes and departure times based on their individual actual/marginal perception to minimize travel time/cost respectively, while the environmental impacts are typically not considered in these principles. The travel choice principle can be formulated as either a nonlinear complementarity problem, variational inequality problem, or fixed-point problem[36].

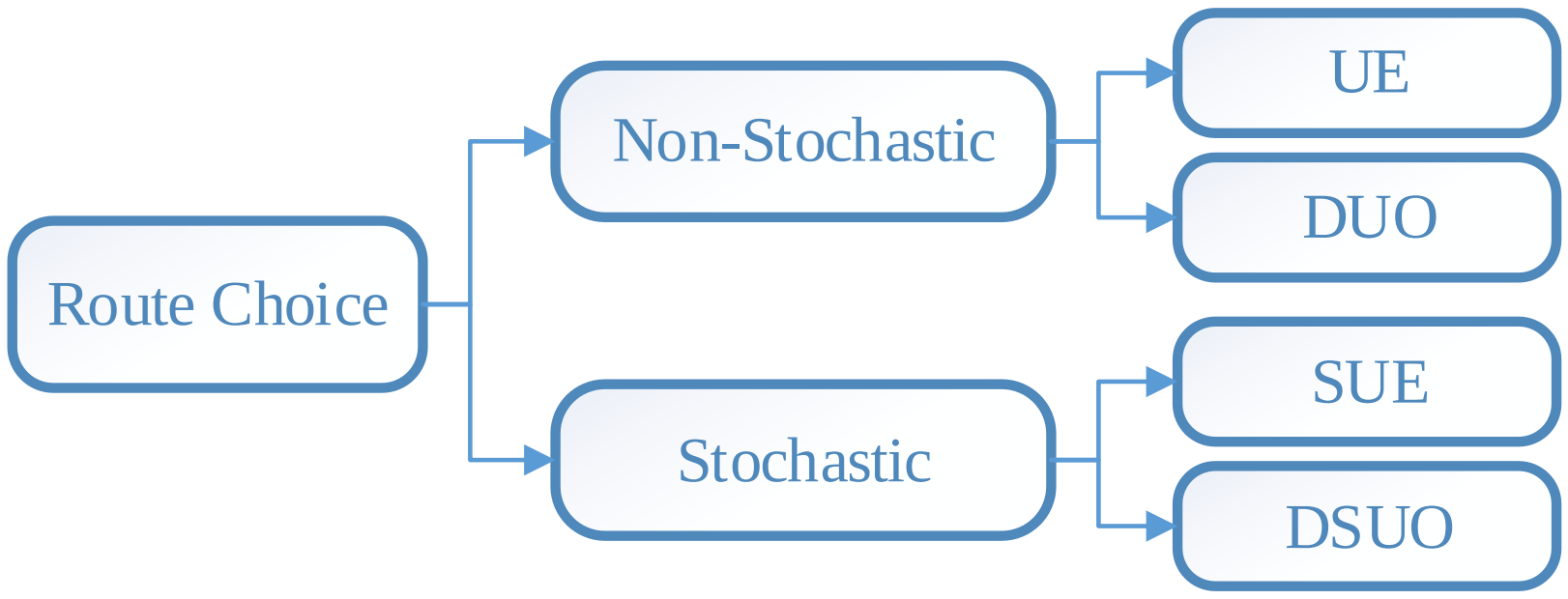

**Fig. 3:** Route Choice Model

### 1-3-4. Organization and conclusion of the traffic assignment approach

With regard to the proposed topics, the steps to implementing the traffic assignment approach can be as follows (**Fig. 4**):

1- Using traffic data, user's travels are determined on the network.
2- The extent of travel between these points is predicted (flow of links is forecasted).
3- The ODM of the network is identified.
4- The routes that users will choose to travel in the network are estimated.
5- Traffic urban network facilities are set up to improve the level of services to users.

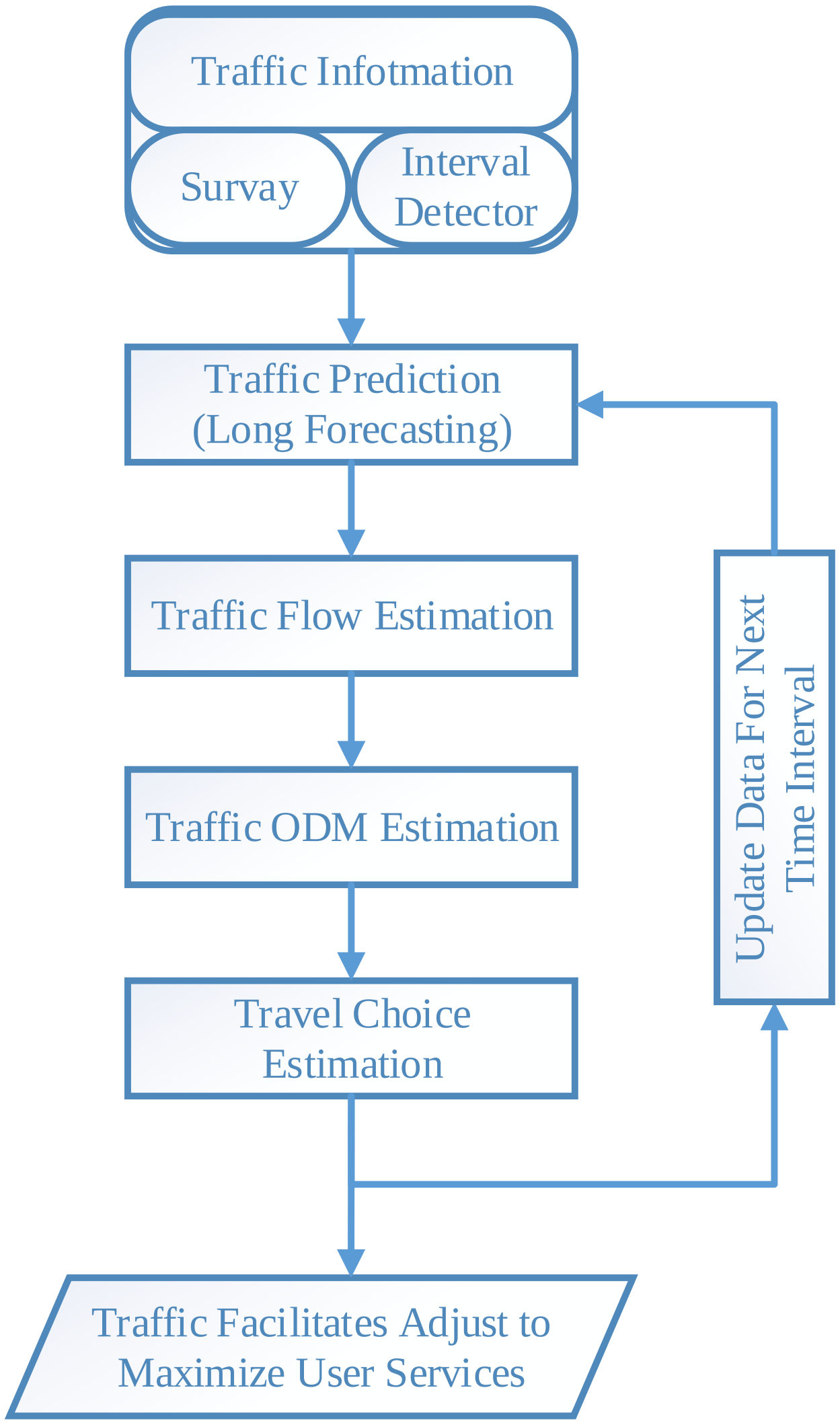

**Fig. 4:** Traffic Assignment Organization

In summary, in this approach, the UTN engineers estimate areas as the origin/destination. Then, they try to identify the routes between different origins/destinations with different statistical mathematical methods and offer the best route to the user for the various periods in a day. In parallel, they adjust different sections of the UTN (e.g. traffic signals) in order to allow users arrive more easily to their destination.

So network users gain a good understanding of the UTN condition and based on that knowledge they choose their route. The advantage of this approach is that regardless of the subsystems in the network, it attempts to optimize the overall network. However, the disadvantage of this approach lies in its uncertainty. Due to the high dependence of its algorithms on statistics and probabilities, one cannot say that the proposed route is definitely the best route. Also, there is no direct control in this approach, and the results are completely dependent on the users' utilization of that.

## 1-4. Traffic Optimization

The growing trend of urbanization and traffic congestion has developed an urgent need to operate our transportation systems with maximum efficiency. One of the most cost-effective methods for dealing with this problem is UTN control/optimization. The UTN is a dynamic system and has an uncertain nature with interdependent subsystems, nonlinearities, and a great number of variables including the vehicle flows, vehicle queues, and semaphore phase times [61]. Due to these complexities, it is necessary to develop an intelligent and economical solution to improve the quality of services for road users. A relatively inexpensive way of alleviating the problem is to ensure optimal use of the existing UTN. With regard to network infrastructure, UTN can be divided into road networks and freeway networks [62].

### 1-4-1. Road Networks

The control tools on the road network can be divided into two categories of traffic signals and traffic signs (e.g. mandatory signs and prohibitory signs). Traffic signal control (TSC) is commonly thought as the most important and effective method for improving the road traffic network efficiency [62]. In particular, the traffic signal setting is performed in two successive steps: single junction signal setting and network coordination [63].

#### 1-4-1-1. Traffic Signal Control

These systems can optimize traffic flows efficiently through two different tools: signal timing and phasing. Signal timing means determining the duration of traffic signal parameters[64]. In general, signal timing plans have three important output parameters which are cycle length, splits, and offsets [65] (**Fig. 5**). Each cycle corresponds to one complete rotation through all of the indications provided at the intersection. Split refers to the total time allocated to each phase in a cycle. The offset is the time lag between the start of green time for successive intersections, which is required to ensure a free flow of vehicles with minimum waiting time along with a specific direction. Traffic signals control the movement of traffic by adjusting the split of each phase assigned within a total cycle time and by modifying the offset [66]. In signal timing approach, the definitions, combinations, and sequencing of stages of the cycle-structures are generated automatically (a stage is a group of non-conflicting phases moving simultaneously) [67].

In parallel, compatible turning movements can be offered with the same traffic signal indications through phasing control. Group-based traffic signal control is the main method of phasing control approaches frequently used for traffic signal systems in many European countries. In this method, the signal setting plan is defined by determining the starting time and duration of the green sign of traffic signals. The main feature of this approach is that the phase sequence is not predetermined. In other words, the group-based method dynamically combines compatible turning movements into phases [64].

According to the literature, the approaches used for signal timing and phasing can be divided into three groups: Pre-Timed control, Actuated control (these control methods are also called adaptive control or traffic responsive), and Semi-Actuated control [65].

### 1- Pre-Timed control

The most basic type of traffic signal control is pre-timed control, where traffic signal timing at an intersection is predetermined and optimized offline based on historical traffic data [68]. This approach has several advantages. Firstly, coordinating the intersections to adjacent junctions is simpler, as signal signs are predictable. Secondly, this approach does not require implementing any detectors, so this approach is very useful under congestion conditions and when detectors fail. Finally, it requires minimal training to set up and maintain. Unfortunately, pre-timed control cannot compensate for unplanned fluctuations in traffic flows, and it tends to be inefficient at isolated intersections where traffic arrivals are random [69].

### 2- Actuated control

Actuated control came into practice with the help of sensing technologies. The idea is that any traffic control action is made under a certain control strategy according to real-time traffic data [62]. These systems have been the most commonly used systems since the early seventies and have shown to be a suitable approach to alleviate traffic congestion as opposed to pre-timed control for signalized intersections [61]. With the introduction of the advanced computer, control, and communication technologies in traffic networks, signal control systems are now able to receive more network-related information and respond in a more congestion-adaptive manner. It should be noted that by increasing the available information, the complexity of algorithms for designing signal timing plans also grows [70].

Different methods have been proposed to control the traffic signals of urban networks, including those based on the optimal control theory and others on artificial intelligence techniques. **Table 1** presents some of these methods.

### 3- Semi-Actuated control

Alongside a fully actuated controller, which uses detector information to choose the timings of all signal groups, and a pre-timed controller, which uses no detector information, it is also possible to have a compromise between these two: the semi-actuated controller [71]. The semi-actuated control uses detector only for the minor road at the intersection. Its main advantage is that it can be used effectively in a coordinated signal system. Also, compared with pre-timed control, it reduces the delay caused by the main road during light traffic periods. Finally, the main road does not require detector for scheduling, and therefore its performance is not at risk in case of failure of these detectors. Semi-actuated control may also be suitable for isolated intersections with a low-speed major road and lighter crossroad volume [69].

## 1-4-1-2. Traffic Signal Coordination

Only by creating the best type of cycle length, split duration, offset, and phase sequence for each intersection of the network, one cannot achieve the maximum efficiency of the road traffic network. The reason is that the interactions of the intersections on each other have a significant effect on traffic network performance. In other words, in the road network, the distance between adjacent intersections is usually short enough that their operation affects each other. As the signal at the upstream intersection turns green, a platoon of vehicles moves to the intersection. When the platoon arrives, if the signal of the corresponding route at the intersection is green, the total delay time and vehicles stopping decrease significantly and the intersection efficiency increase. To this aim, intersections must be coordinated together.

Coordinating models can be divided into the tree structure and the cyclic structure. In the tree structure, an intersection is considered as a major node, and the other intersections are coordinated with it in a hierarchical manner. One-directional and two- directional green wave model are two types of the most implemented tree structure intersection coordination. For a green wave, the traffic signals over several intersections are coordinated for one main direction, allowing continuous traffic flow (without stopping) over these intersections for this main direction. This allows a platoon of vehicles to flow through the network with as minimum hindrance as possible [71]. In the cyclic structure, based on the density and capacity of each road connected to the intersection, each direction has its percentage of priority. Based on these priority percentage and distance of the adjacent junction, the cycle length, split duration, and offset are determined (**Fig. 5**). Some of the coordination methods are provided in **Table 1**.

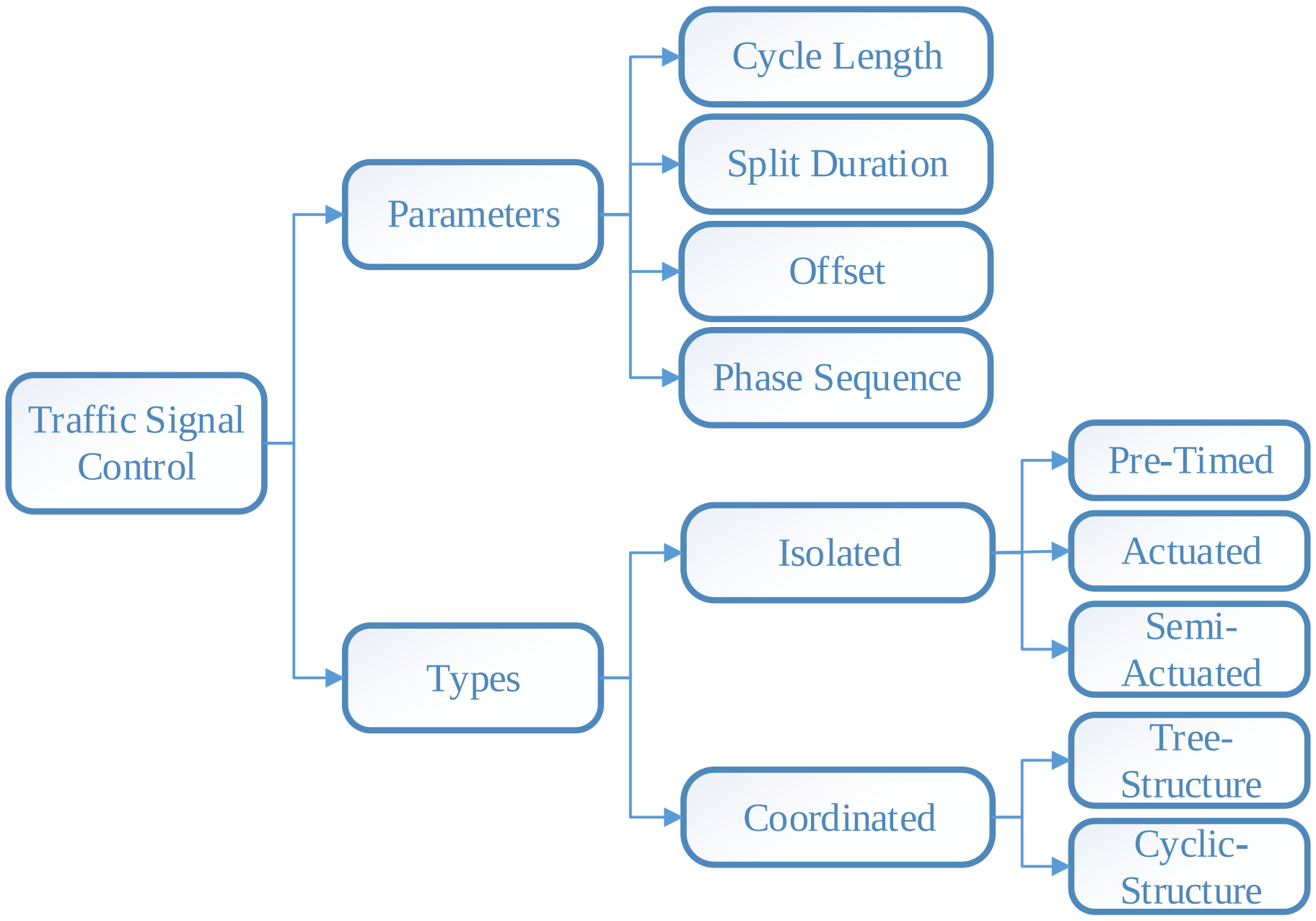

**Fig. 5:** Traffic Signal Control

It should be noted that the aim of coordination in congestion and noncongestion manner is different. In congestion manner, coordination tries to prevent traffic jam in upstream or downstream junction by stopping the vehicle in that direction, while in noncongestion manner, the optimization function tries to minimize total delay time by maximizing flows.

**Tab. 1:** Different methods of junction control and coordination

| Reference number | Date of publishing | Fuzzy logic systems | Multi-agent systems | Artificial neural network | Reinforcement learning | Evolutionary Computation | Swarm Intelligence | Game Theory | Simulation | Implementation | Isolated junction | Traffic network |
|---|---|---|---|---|---|---|---|---|---|---|---|---|
| [72] | 1993 | | | ✓ | | | | | ✓ | | | ✓ |
| [73] | 1999 | ✓ | | | | | | | | ✓ | ✓ | |
| [74] | 1999 | ✓ | | | | | | | ✓ | | | ✓ |
| [75] | 2000 | | ✓ | | | | | | ✓ | | | ✓ |
| [76] | 2001 | ✓ | | | ✓ | | | | ✓ | | ✓ | |
| [77] | 2004 | | ✓ | | ✓ | | | | | ✓ | ✓ | |
| [78] | 2005 | | ✓ | | | | | ✓ | | | ✓ | |
| [79] | 2005 | ✓ | | | ✓ | | | | ✓ | | | ✓ |
| [80] | 2006 | | | ✓ | ✓ | | | | | | | ✓ |
| [81] | 2009 | ✓ | | ✓ | | | | | ✓ | | | ✓ |
| [82] | 2009 | | | | ✓ | | | | ✓ | | | ✓ |
| [83] | 2010 | | ✓ | | | | | | | ✓ | | ✓ |
| [84] | 2010 | | | | | | | ✓ | ✓ | | ✓ | |
| [85] | 2011 | ✓ | | | | ✓ | | | ✓ | | ✓ | |
| [86] | 2011 | ✓ | ✓ | | | | | | ✓ | | | ✓ |
| [87] | 2011 | | | | ✓ | | ✓ | | ✓ | | | ✓ |
| [88] | 2013 | | ✓ | | ✓ | | | | ✓ | | | ✓ |
| [89] | 2013 | | | | | ✓ | | | ✓ | | | ✓ |
| [90] | 2014 | ✓ | ✓ | | | ✓ | | | ✓ | | | ✓ |
| [91] | 2015 | | ✓ | ✓ | | | | | | ✓ | ✓ | |
| [92] | 2015 | | ✓ | | | | ✓ | | ✓ | | | ✓ |
| [93] | 2015 | | | ✓ | | | ✓ | | ✓ | | ✓ | |
| [94] | 2015 | ✓ | | | ✓ | | | | ✓ | | ✓ | |
| [95] | 2015 | | | | | | | ✓ | ✓ | | ✓ | |
| [96] | 2016 | | ✓ | | ✓ | | | | ✓ | | | ✓ |
| [65] | 2017 | | | | | | ✓ | | ✓ | | | ✓ |
| [68] | 2017 | | | ✓ | ✓ | | | | ✓ | | ✓ | |
| Sum | 27 | 9 | 10 | 6 | 10 | 3 | 4 | 3 | 21 | 4 | 11 | 16 |

According to the **Table 1**, it can be concluded that most researchers have used fuzzy logic systems, multi-agent systems, and artificial neural networks as the basic methods for controlling and optimizing the intersections. In recent years, they have combined other optimizing methods with these to determine the intersection parameters (**Fig. 6**).

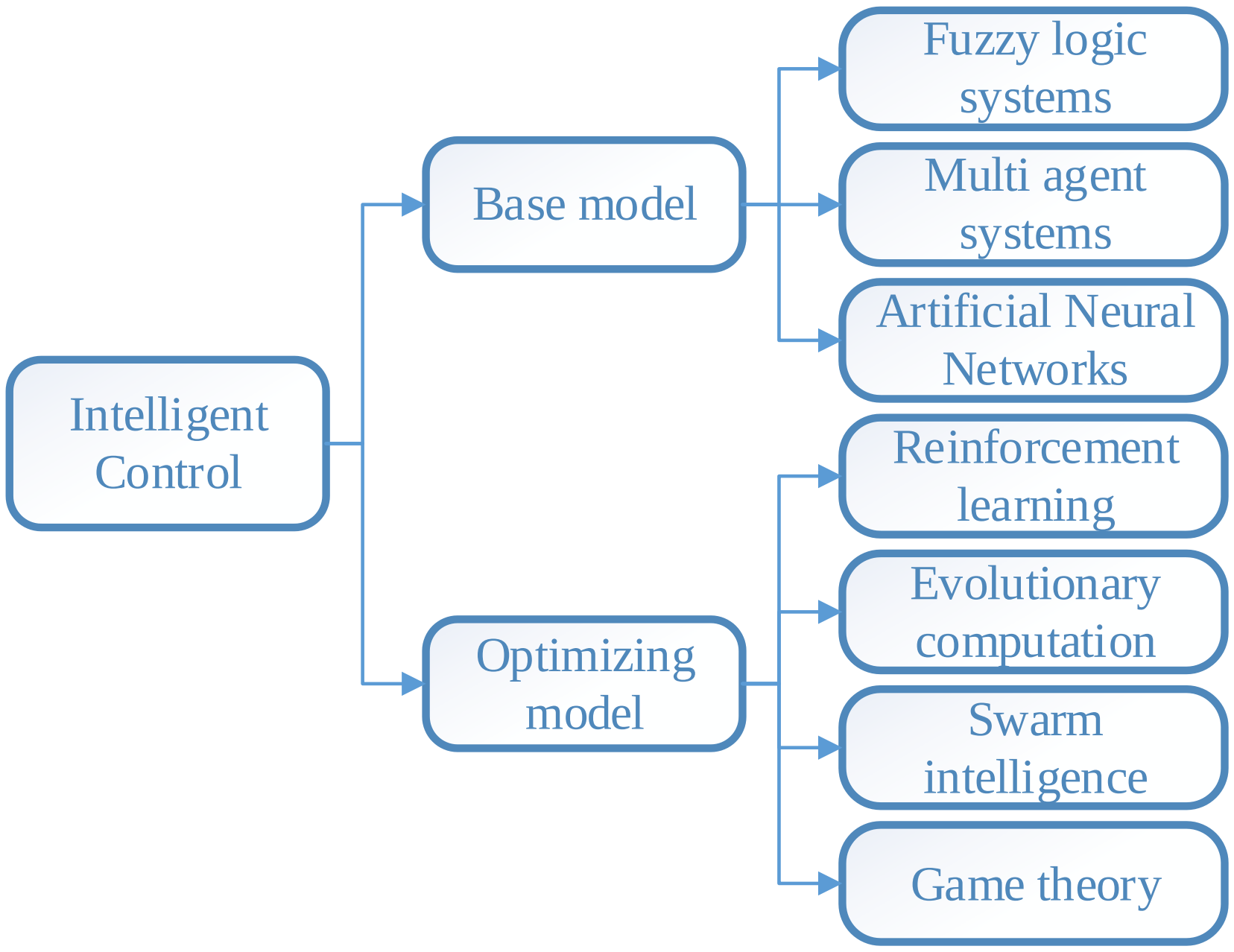

**Fig. 6:** Intersection control methods

### 1-4-2. Freeway networks

Freeway networks had been originally conceived to provide almost unlimited mobility without traffic signals interruptions for UTN users. The rapid rise in the demand has resulted in the creation of severe congestions, both recurrent (occurring daily during rush hours) and non- recurrent (due to accidents). In other words, congestion on freeways develops when too many vehicles try to use them without being controlled. When congestion develops, the outflow from the loading area is reduced, and the off-ramps and pathways filled with congestion are blocked, which may even lead to gridlocks in some cases. Prevention or reduction of traffic congestion on freeway networks may significantly improve the infrastructure efficiency in terms of total time spent. To this aim, engineers can act in two ways [97]:

- Decreasing demand by ramp metering (RM) and driver informing
- Mainstream control by lane control, variable speed limits, congestion warning, etc.

Using traffic signals, RM decouples platoons into individual vehicles, while variable speed limits (VSLs) intend to keep platoons intact [98].

Freeway traffic control by means of VSL aims at homogenizing speed (regulating drivers' behavior) and preventing traffic breakdown. Homogenizing speed reduces the speed differences between vehicles, which can result in a higher (and safer) traffic flow, thus achieving a better utilization of the space, also causing a higher density and flow. In this approach, the speed limits are close to but above the critical speed (the speed that results in the maximum flow). The traffic breakdown prevention approach focuses more on preventing too high densities, and also tries to keep the speed lower than the critical point. Preventing the shock wave and avoiding postpone congestion are two strategies that use this approach [99].

### 1-4-2-1. Ramp Meter

RM is the most direct and efficient way to control freeway traffic. It is defined as a method of improving overall freeway operations by limiting, regulating, and timing the entrance of vehicles from one or more ramps onto the freeway [100]. Ramp meters are traffic signals that control traffic at entrances to freeways. They can be used for two different purposes. Firstly, when traffic is dense, in order to keep the density below the critical value, RM can prevent the traffic breakdown by adjusting the metering rate. It also encourages drivers to bypass congestion on a freeway using the road network. In other words, It discourages them to use the freeway with increasing travel time [101].

So far, different types of RM have been proposed. Because of their similar operation to that of the traffic signal, they can be divided into three groups based on their signaling time model: fixed time, pre-timed, and traffic responsive.

#### 1- Fixed time

Fixed time metering rates are below the non-metered level. The simplest control policy, brought from the traffic signal control, is based on the fixed time metering plans. The idea behind this operation is to break up vehicle platoons before entering the freeway. This can be potentially beneficial in the reduction of accidents by providing a smooth merging of the mainstream and the on-ramp flows. This method is usually implemented only under light traffic congestion as an initial operating strategy [102].

#### 2- Pre-timed

One of the pioneer experiments of RM control strategy is pre-timed control. The metering rate is determined by analyzing historical volume-capacity conditions during a specific period of time for the target ramp. The advantage of this approach lies in its simplicity and implement ability as an initial operating strategy until individual ramps can be incorporated into a traffic responsive system. The disadvantage of this approach, however, is its lack of response to momentary traffic conditions: either change in demands or capacities [103].

#### 3- Traffic Responsive

Traffic-responsive ramp-metering strategies, as opposed to fixed-time strategies, are based on real-time measurements from sensors mounted in the freeway network. After a relevant processing, this information is given to the controllers which calculate control decisions based on their heuristic or optimal policies. These decisions are then translated into traffic signals and sent to the ramp meters. The full setup, among many other hardware and software instruments, contains repeaters, access points, transmitters, data collection points, state estimators, and flow predictors [102].

Two kinds of grouping can be used for classifying traffic-responsive ramp-metering. The first is coordination position and the second is coordination aim. Based on the coordination position, these ramps can be classified as coordinated with either a local congestion, upstream freeway congestion, downstream freeway congestion, or with a connected road network or as a combination of these. Based on the desired target, they can be divided into demand/capacity, or its derivative called occupancy rate or both of them. In other words, in order to solve freeway congestion, engineers should decrease the demand or increase the occupancy or consider both of them in a parallel and coordinated manner. Based on these groups, such methods are proposed in **Table 2**.

**Tab. 2:** RM Algorithm

| | Demand/Capacity | Density Control | Both Demand and Density |
|---|---|---|---|
| Local Congestion | Demand/Capacity Control[104] | Occupancy-rate Control[105] | ALINEA[1][106] |
| Upstream Congestion | | Bottleneck Algorithm[107] | |
| Downstream Congestion | Helper Algorithm[103] | | |
| Upstream Arterial Network | Sperry Algorithm[103] | | |
| Network Coordination | Zone Algorithm[108] | ARMS[2][109] | Compass Algorithm, METALINE, HERO[110] |

As with traffic junction control algorithms, so far engineers have presented different algorithms for modeling and optimizing the mathematical function of traffic-responsive ramp-metering. Lyapunov Stability[111], Linear Quadratic Control[112], Fuzzy Logic Algorithm[113], [114], Artificial Neural Network[115], [116], Reinforcement learning Algorithm[117], [118] and Genetic Algorithm[119], [120] are such methods that are used for traffic-responsive ramp-metering.

**Fig. 7** displays different freeway network control approaches.

---

[1] A Local Feedback Control Law for On-Ramp Metering
[2] Advanced Real-time Metering System

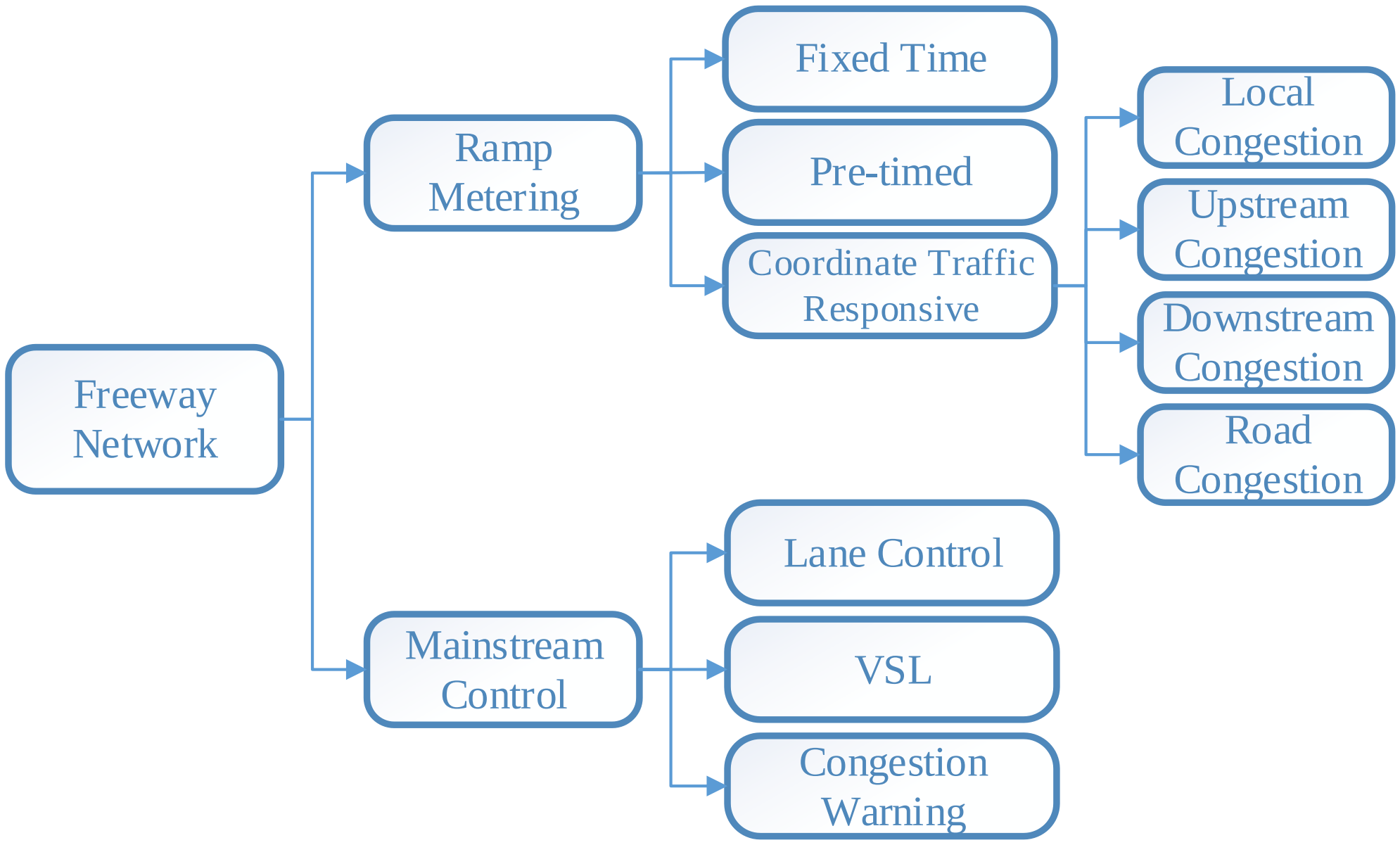

**Fig. 7:** Freeway Network

### 1-4-3. Organization and conclusion of traffic optimization approach

With regard to the proposed topics, the steps to implementing the traffic optimization approach can be stated as follows (**Fig. 8**):

1- Using traffic data, the user's demand in each route of an intersection is determined.
2- The cycle length, split duration, offset and phase sequence are optimized in order to maximize the flow through that.
3- Traffic condition is predicted for the adjacent intersection.
4- The intersection becomes coordinated to adjacent nodes given the predicted information.

In summary, in this approach, the UTN engineers first optimize each node individually. They then try to coordinate each intersection to other nodes of the network with different computational intelligence and control methods, and then determine the best extent of signal timing and phasing for the various periods in the day. In parallel, they also send these parameters to the management center for traffic assignment applications.

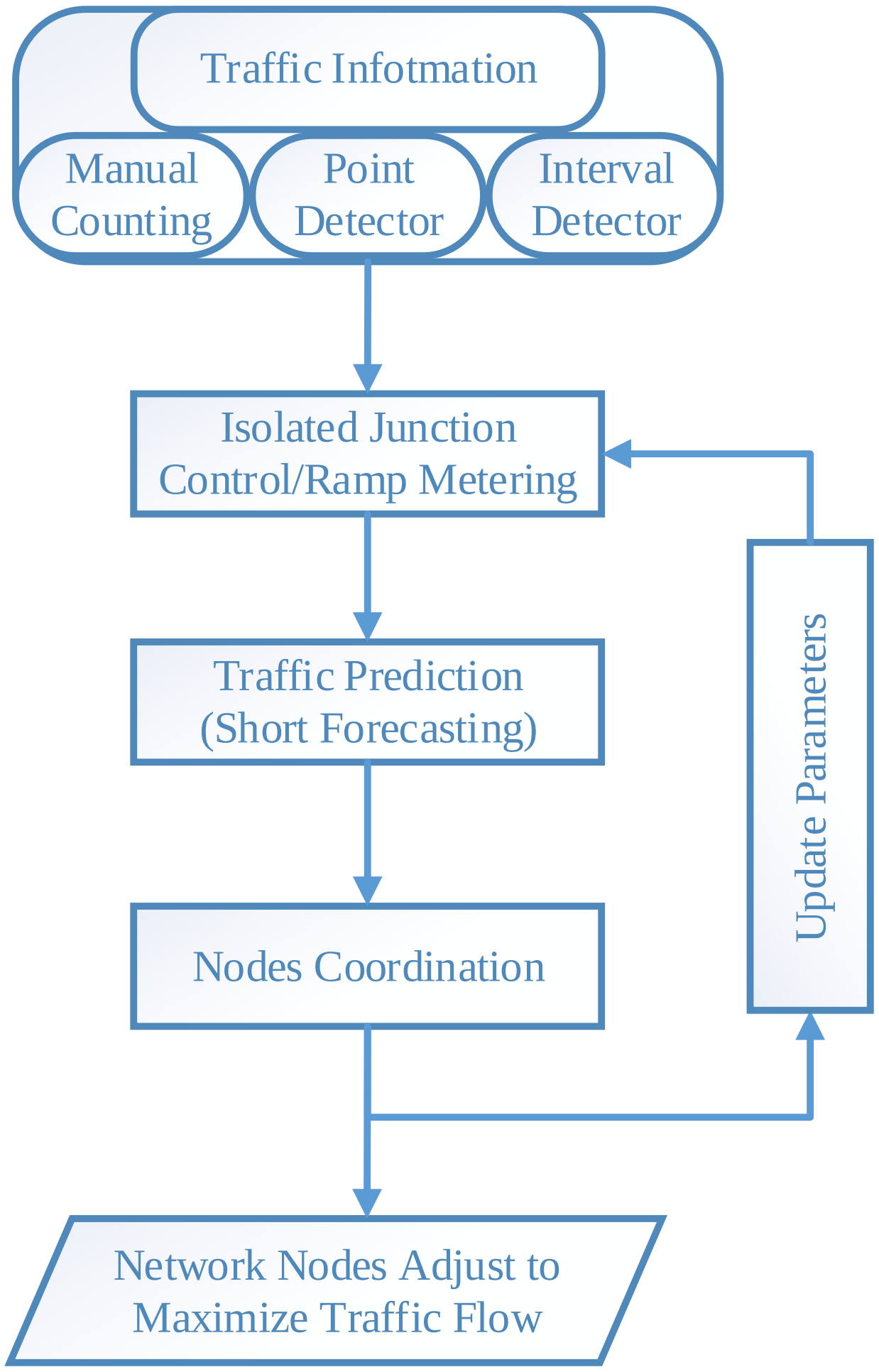

**Fig. 8:** Traffic Optimization Organization

Direct control and gaining a proper understanding of network details to UTN managers are among the benefits of this approach. In this approach, it is assumed that the optimality of the subsystems and their coordination with each other will lead to the optimization of the overall system. Such an assumption is not true for nonlinear systems, and it is the main disadvantage of this method. In other words, due to the different capacities of different network areas and the mismatch of their demand to their capacity, some areas of the network are high density, while some others are low density. Indeed, optimizing the condition of the high-density areas and low-density areas separately and coordinating them with each other does not mean that the overall network is reaching its optimal status. Further, due to the network's size, lack of the hardware potential, and dynamic features of UTN, the whole network cannot be placed under a centralized control and then be generally optimized. Note that dividing network into smaller sectors is essential for control. So this approach, alone, cannot optimize the overall network. For optimizing the overall network, traffic optimization approach must be combined with the traffic assignment approach.

## 1-5. Traffic Prediction

Forecasting traffic flows is an essential requirement for ATMS. This section provides the necessary data for traffic assignment and traffic optimization. In other words, ATMS for evaluating the condition of UTN in real time collect traffic information, and then by applying traffic forecasting techniques, provides the optimal strategy for improving traffic network flows and guiding network users. Indeed, traffic forecasting can be used effectively to help network users in route choice. It can also be employed in traffic control solutions in order to prevent traffic congestions. Variables, whose prediction is used, are:

1. Flow
2. Density or occupancy (when the detector is occupied by vehicles)
3. Travel time on a link or on a route
4. The average speed of all cars that pass through a given period of time

The accuracy of prediction is expressed in terms of probabilities. Root mean square error (RMSE) and mean absolute percentage error (MAPE) are the most commonly used measures. RMSE expresses the expected error value in a unit the same as the unit of data, while the MAPE expresses the error as a percentage, making it possible to understand the error rate.

In recent years, various methods have been used to predict the traffic condition. The studies in this field can be divided into three main trends:

1- Naive methods
2- Parametric methods
3- Nonparametric methods

### 1-5-1. Naive methods

In these methods, no traffic flow model is considered, so they have little computation and are easy to implement. The low accuracy of these methods is their biggest disadvantage.

**1- Instantaneous**

When instantaneous traffic conditions are used to predict traffic, it is assumed that the traffic situation will remain constant infinitely. Although this method is very fast and simple, due to the lack of traffic condition stability, its accuracy is very poor [121].

**2- Historical Average**

The average of past traffic data generates a historical average of traffic variables. Compared to advanced techniques, the historical average is not superior at all [122], [123].

**3- The combination of the instantaneous and the historical average methods**

The combination of the instantaneous measurement and the average of past data has been used in some papers in various ways for prediction purposes [124].

**4- Clustering**

The clustering methods separate the same traffic pattern on each day of the week and average the traffic variables in each of the patterns individually. Sometimes clustering is also used as a preprocessor of information [125], [126].

### 1-5-2. Parametric methods

The famous model that is presented based on parametric methods is called the Auto-Regressive Integrated Moving Average (ARIMA) model. This method is a complete solution for most time series issues. In other words, ARIMA is a prediction technique that determines the future values of a time series data based on their internal information. For the first time, this method was proposed by Box and Jenkin in 1973[127], so in some papers, this method is called Box-Jenkin. The main application of this technique is for short-term prediction. This method works best when the data have a stable behavior or show a specific time pattern, In order to predict traffic condition, this method was used for the first time in 1979 [128]. Since then, the ARIMA model has been used as the main method for analyzing the time series traffic data, especially in the area of traffic forecasting.

The first step in using the ARIMA approach is to investigate the data stability. Data stability means that their collection remains constant over time. In most economic and commercial data, a steady trend is observed in data changes. Stability does not exist in this category of data. On the other hand, the data have a steady deviation in their changes over time. Without observing stability conditions in the data, the ARIMA method will not respond appropriately. In this case, deriving from data is an excellent solution for converting them to stable data. This is done by reducing the current measurement from the previous one. If this conversion is applied to the data only once, the generated data is named the first-order derivative. This stability will occur in data that grows at an almost constant rate. If the data grow at an incremental rate, one can use a similar process and derive from the data again. In this case, the generated data are called second-order derivatives [129].

### 1-5-3. Nonparametric methods

In the case of non-parametric models, the most widely used method is the artificial neural network (ANN). Multilayer Perceptron (MLP)[130], Back-Propagation Neural Network (BPNN) [131], [132], and Radial Basis Function Neural Network (RBFNN)[133] are the most common

ANN models used to predict short-term traffic flow. Compared to BPNN, RBFNN requires shorter training time and provides better performance.

The goal of all these studies is to obtain more accurate prediction results within a shorter time.

## 1-6. Conclusion and Development Strategy

Based on what has been explained here, traffic assignment must be combined with traffic optimization for giving the best service to network users. Also, traffic optimization should be mixed with traffic assignment to maximize the overall network efficiency. So ATMS can be organized as follows (**Fig. 9**):

1- Using traffic information to control traffic nodes.
2- Determining parameters of nodes (such as cycle length and split duration for signalized junctions, and duty cycle for RM) to maximize efficiency.
3- Predicting traffic conditions for adjacent nodes.
4- Coordinating nodes together.
5- Predicting the percentage of network users distributed across the network.
6- Estimating the flow of network link.
7- Estimating the network ODM in a hierarchical manner.
8- Estimating users' network travel.
9- Updating traffic information for the next time interval.

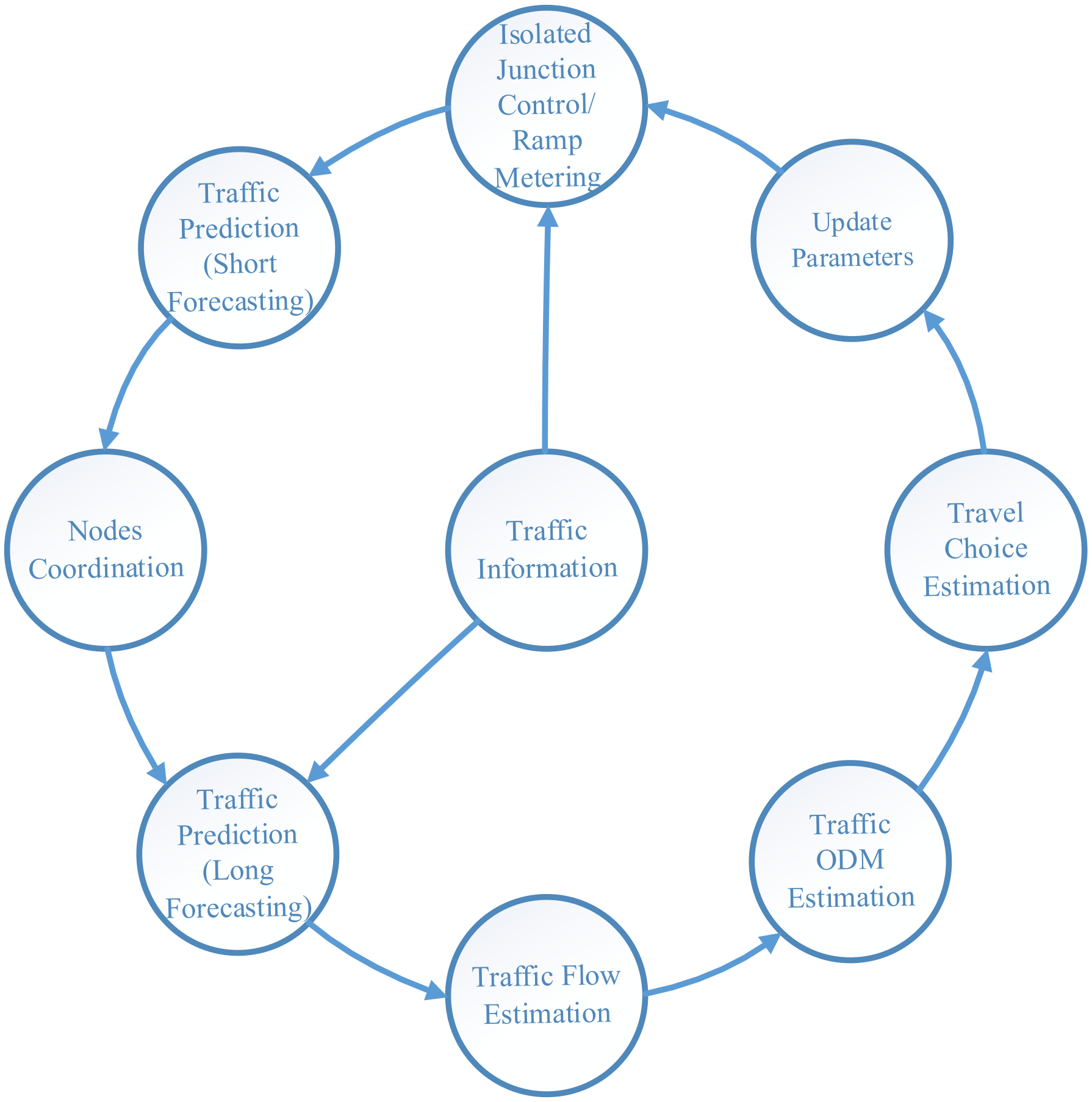

**Fig. 9:** ATMS development strategy

[1] P. Cramton, R. R. Geddes, and A. Ockenfels, "Set road charges in real time to ease traffic," *Nature*, vol. 560, no. 7716, p. 23, Aug. 2018.

[2] "$CO_2$ and other Greenhouse Gas Emissions," *Our World in Data*. [Online]. Available: https://ourworldindata.org/co2-and-other-greenhouse-gas-emissions. [Accessed: 06-Aug-2018].

[3] "Air Pollution," *Our World in Data*. [Online]. Available: https://ourworldindata.org/air-pollution. [Accessed: 05-Aug-2018].

[4] "Vehicles, Air Pollution, and Human Health," *Union of Concerned Scientists*. [Online]. Available: https://www.ucsusa.org/clean-vehicles/vehicles-air-pollution-and-human-health. [Accessed: 05-Aug-2018].

[5] INRIX, "INRIX Global Traffic Scorecard," *INRIX - INRIX*. [Online]. Available: http://inrix.com/scorecard/. [Accessed: 05-Aug-2018].

[6] "The hidden cost of congestion," *The Economist*, 28-Feb-2018.

[7] R. P. Roess, E. S. Prassas, and W. R. McShane, *Traffic Engineering*, 4 edition. Upper Saddle River, NJ: Prentice Hall, 2010.

[8] M. Smith, "Traffic signal control and route choice: A new assignment and control model which designs signal timings," *Transp. Res. Part C Emerg. Technol.*, vol. 58, pp. 451–473, Sep. 2015.

[9] L. Du, S. Peeta, and Y. H. Kim, "An adaptive information fusion model to predict the short-term link travel time distribution in dynamic traffic networks," *Transp. Res. Part B Methodol.*, vol. 46, no. 1, pp. 235–252, Jan. 2012.

[10] D. Smith, *Handbook of Simplified Practice for Traffic Studies*. Center for Transportation Research and Education Iowa State University, 2002.

[11] A. Nkaro, *Traffi c Data Collection and Analysis*, vol. 54. Gaborone, Botswana: Ministry of Works and Transport Roads Department, 2003.

[12] F. Calabrese, G. D. Lorenzo, L. Liu, and C. Ratti, "Estimating Origin-Destination Flows Using Mobile Phone Location Data," *IEEE Pervasive Comput.*, vol. 10, no. 4, pp. 36–44, Apr. 2011.

[13] M. Grosh and P. Glewwe, *Designing Household Survey Questionnaires for Developing Countries*. The World Bank, 2000.

[14] J.-P. Hubert, J. Armoogum, K. W. Axhausen, and J.-L. Madre, "Immobility and Mobility Seen Through Trip-Based Versus Time-Use Surveys," *Transp. Rev.*, vol. 28, no. 5, pp. 641–658, Sep. 2008.

[15] R. B. Voas, J. Wells, D. Lestina, A. Williams, and M. Greene, "Drinking and driving in the United States: The 1996 national roadside survey," *Accid. Anal. Prev.*, vol. 30, no. 2, pp. 267–275, Mar. 1998.

[16] H. Yang, C. Yang, and L. Gan, "Models and algorithms for the screen line-based traffic-counting location problems," *Comput. Oper. Res.*, vol. 33, no. 3, pp. 836–858, Mar. 2006.

[17] P. Bonnel, E. Hombourger, A.-M. Olteanu-Raimond, and Z. Smoreda, "Passive Mobile Phone Dataset to Construct Origin-destination Matrix: Potentials and Limitations," *Transp. Res. Procedia*, vol. 11, pp. 381–398, Jan. 2015.

[18] U. Mori, A. Mendiburu, M. Álvarez, and J. A. Lozano, "A review of travel time estimation and forecasting for Advanced Traveller Information Systems," *Transp. Transp. Sci.*, vol. 11, no. 2, pp. 119–157, Feb. 2015.

[19] G. Leduc, "Road Traffic Data: Collection Methods and Applications," 2008.

[20] L. Gong, T. Morikawa, T. Yamamoto, and H. Sato, "Deriving Personal Trip Data from GPS Data: A Literature Review on the Existing Methodologies," *Procedia - Soc. Behav. Sci.*, vol. 138, pp. 557–565, Jul. 2014.

[21] Y.-S. Lee and S.-B. Cho, "Activity recognition with android phone using mixture-of-experts co-trained with labeled and unlabeled data," *Neurocomputing*, vol. 126, pp. 106–115, Feb. 2014.

[22] A. Moiseeva and H. Timmermans, “Imputing relevant information from multi-day GPS tracers for retail planning and management using data fusion and context-sensitive learning,” *J. Retail. Consum. Serv.*, vol. 17, no. 3, pp. 189–199, May 2010.

[23] F. Calabrese, M. Diao, G. Di Lorenzo, J. Ferreira, and C. Ratti, “Understanding individual mobility patterns from urban sensing data: A mobile phone trace example,” *Transp. Res. Part C Emerg. Technol.*, vol. 26, pp. 301–313, Jan. 2013.

[24] “Uber,” *Wikipedia*. 02-Apr-2018.

[25] S. Guo, Y. Liu, K. Xu, and D. M. Chiu, “Understanding passenger reaction to dynamic prices in ride-on-demand service,” in *2017 IEEE International Conference on Pervasive Computing and Communications Workshops (PerCom Workshops)*, 2017, pp. 42–45.

[26] B. Martin and P. Scott, “Automatic Vehicle Identification: A Test of Theories of Technology,” *Sci. Technol. Hum. Values*, vol. 17, no. 4, pp. 485–505, 1992.

[27] A. Bhavke and S. Pai, “Advance automatic toll collection vehicle detection during collision using RFID,” in *2017 International Conference on Nascent Technologies in Engineering (ICNTE)*, 2017, pp. 1–5.

[28] J. H. Rillings and R. J. Betsold, “Advanced driver information systems,” *IEEE Trans. Veh. Technol.*, vol. 40, no. 1, pp. 31–40, Feb. 1991.

[29] F. M. Oliveira-Neto, L. D. Han, and M. K. Jeong, “An Online Self-Learning Algorithm for License Plate Matching,” *IEEE Trans. Intell. Transp. Syst.*, vol. 14, no. 4, pp. 1806–1816, Dec. 2013.

[30] Y. Malinovskiy, Y.-J. Wu, Y. Wang, and U. K. Lee, “Field Experiments on Bluetooth-Based Travel Time Data Collection,” presented at the Transportation Research Board 89th Annual MeetingTransportation Research Board, 2010.

[31] M. Abbott-Jard, H. Shah, and A. Bhaskar, “Empirical evaluation of Bluetooth and Wifi scanning for road transport,” presented at the Australasian Transport Research Forum (ATRF), 36th, 2013, Brisbane, Queensland, Australia, 2013.

[32] J. Barceló, L. Montero, L. Marqués, and C. Carmona, “Travel Time Forecasting and Dynamic Origin-Destination Estimation for Freeways Based on Bluetooth Traffic Monitoring,” *Transp. Res. Rec. J. Transp. Res. Board*, vol. 2175, pp. 19–27, Dec. 2010.

[33] S. Bretschneider, *Mathematical Models for Evacuation Planning in Urban Areas*, 1st ed. Chapter 2: Literature Review: Springer-Verlag Berlin Heidelberg, 2013.

[34] E. Mitsakis, J. M. Salanova, and G. Giannopoulos, “Combined dynamic traffic assignment and urban traffic control models,” *Procedia - Soc. Behav. Sci.*, vol. 20, pp. 427–436, Jan. 2011.

[35] I. BOMBAY, “Transport Engineering,” *NPTEL*, 2009. [Online]. Available: http://nptel.ac.in/courses/105101087/. [Accessed: 19-Apr-2018].

[36] Y. Wang, W. Y. Szeto, K. Han, and T. L. Friesz, “Dynamic traffic assignment: A review of the methodological advances for environmentally sustainable road transportation applications,” *Transp. Res. Part B Methodol.*, vol. 25, pp. 1–25, 2018.

[37] W. Y. Szeto and H. K. Lo, “Dynamic Traffic Assignment: Properties and Extensions,” *Transportmetrica*, vol. 2, no. 1, pp. 31–52, Jan. 2006.

[38] S. Bera and K. V. K. Rao, “Estimation of origin-destination matrix from traffic counts: the state of the art,” *Eur. Transp. Trasp. Eur.*, no. 49, pp. 2–23, 2011.

[39] R. Frederix, F. Viti, R. Corthout, and C. Tampère, “New Gradient Approximation Method for Dynamic Origin-Destination Matrix Estimation on Congested Networks,” *Transp. Res. Rec. J. Transp. Res. Board*, vol. 2263, pp. 19–25, Dec. 2011.

[40] S. Lee, B. Heydecker, Y. H. Kim, and E. Shon, “Dynamic OD estimation using three phase traffic flow theory,” *Journal of Advanced Transportation*.

[41] L. G. Willumsen, *Estimation of an O-D Matrix from Traffic Counts – A Review*. Institute of Transport Studies, University of Leeds, 2017.

[42] M. B. Gonçalves and I. Ulysséa-Neto, “The Development of a New Gravity—Opportunity Model for Trip Distribution:,” *Environ. Plan. A*, Jul. 2016.
[43] A. Wilson, *Entropy in Urban and Regional Modelling*, 1 edition. Routledge, 2011.
[44] A. G. Wilson, “The Use of the Concept of Entropy in System Modelling,” *J. Oper. Res. Soc.*, vol. 21, no. 2, pp. 247–265, Jun. 1970.
[45] H. Yang, T. Sasaki, Y. Iida, and Y. Asakura, “Estimation of origin-destination matrices from link traffic counts on congested networks,” *Transp. Res. Part B Methodol.*, vol. 26, no. 6, pp. 417–434, Dec. 1992.
[46] K. Parry and M. L. Hazelton, “Bayesian inference for day-to-day dynamic traffic models,” *Transp. Res. Part B Methodol.*, vol. 50, pp. 104–115, Apr. 2013.
[47] H. Spiess, “A maximum likelihood model for estimating origin-destination matrices,” *Transp. Res. Part B Methodol.*, vol. 21, no. 5, pp. 395–412, Oct. 1987.
[48] D. P. Watling, “Maximum likelihood estimation of an origin-destination matrix from a partial registration plate survey,” *Transp. Res. Part B Methodol.*, vol. 28, no. 4, pp. 289–314, Aug. 1994.
[49] R. E. Kalman, “A New Approach to Linear Filtering and Prediction Problems,” *J. Basic Eng.*, vol. 82, no. 1, pp. 35–45, Mar. 1960.
[50] M. Bierlaire and F. Crittin, “An Efficient Algorithm for Real-Time Estimation and Prediction of Dynamic OD Tables,” *Oper. Res.*, vol. 52, no. 1, pp. 116–127, Feb. 2004.
[51] A. Ahmed, “Integration of Real-time Traffic State Estimation and Dynamic Traffic Assignment with Applications to Advanced Traveller Information Systems,” phd, University of Leeds, 2015.
[52] C. Gawron, “An Iterative Algorithm to Determine the Dynamic User Equilibrium in a Traffic Simulation Model,” *Int. J. Mod. Phys. C*, vol. 09, no. 03, pp. 393–407, May 1998.
[53] C. G. Chorus, T. A. Arentze, and H. J. P. Timmermans, “A Random Regret-Minimization model of travel choice,” *Transp. Res. Part B Methodol.*, vol. 42, no. 1, pp. 1–18, Jan. 2008.
[54] H. S. Mahmassani, S. Peeta, G.-L. Chang, and T. Junchaya, “Areview of dynamic assignment and traffic simulation models for ADIS/ATMS applications,” Center for transport an on research, the university of Texas at Austin, Technical DTFH61-90-R-00074, 1991.
[55] S. Peeta and H. S. Mahmassani, “System optimal and user equilibrium time-dependent traffic assignment in congested networks,” *Ann. Oper. Res.*, vol. 60, no. 1, pp. 81–113, Dec. 1995.
[56] K. Saw, B. K. Katti, and G. Joshi, “Literature Review of Traffic Assignment: Static and Dynamic,” *Int. J. Transp. Eng.*, vol. 2, no. 4, pp. 339–347, Apr. 2015.
[57] D. K. Merchant and G. L. Nemhauser, “A Model and an Algorithm for the Dynamic Traffic Assignment Problems,” *Transp. Sci.*, vol. 12, no. 3, pp. 183–199, Aug. 1978.
[58] V. Klee and G. Minty, “How good is the simplex algorithm?,” in *Inequalities*, vol. III, O. Shisha, Ed. Academic Press, 1972, pp. 159–175.
[59] T. L. Friesz, J. Luque, R. L. Tobin, and B.-W. Wie, “Dynamic Network Traffic Assignment Considered as a Continuous Time Optimal Control Problem,” *Oper. Res.*, vol. 37, no. 6, pp. 893–901, Dec. 1989.
[60] B. Ran, R. W. Hall, and D. E. Boyce, “A link-based variational inequality model for dynamic departure time/route choice,” *Transp. Res. Part B Methodol.*, vol. 30, no. 1, pp. 31–46, Feb. 1996.
[61] S. Touhbi *et al.*, “Adaptive Traffic Signal Control : Exploring Reward Definition For Reinforcement Learning,” *Procedia Comput. Sci.*, vol. 109, pp. 513–520, Jan. 2017.
[62] D. Zhao, Y. Dai, and Z. Zhang, “Computational Intelligence in Urban Traffic Signal Control: A Survey,” *IEEE Trans. Syst. Man Cybern. Part C Appl. Rev.*, vol. 42, no. 4, pp. 485–494, Jul. 2012.
[63] G. E. Cantarella and A. Sforza, “Network Design Models and Methods for Urban Traffic Management,” in *Urban Traffic Networks*, Springer, Berlin, Heidelberg, 1995, pp. 123–153.
[64] J. Jin and X. Ma, “A group-based traffic signal control with adaptive learning ability,” *Eng. Appl. Artif. Intell.*, vol. 65, pp. 282–293, Oct. 2017.

[65] H. S. E. Chuo, M. K. Tan, A. C. H. Chong, R. K. Y. Chin, and K. T. K. Teo, “Evolvable traffic signal control for intersection congestion alleviation with enhanced particle swarm optimisation,” in *2017 IEEE 2nd International Conference on Automatic Control and Intelligent Systems (I2CACIS)*, 2017, pp. 92–97.
[66] P. G. Balaji and D. Srinivasan, “Multi-Agent System in Urban Traffic Signal Control,” *IEEE Comput. Intell. Mag.*, vol. 5, no. 4, pp. 43–51, Nov. 2010.
[67] S. Lee, S. C. Wong, and P. Varaiya, “Group-based hierarchical adaptive traffic-signal control part I: Formulation,” *Transp. Res. Part B Methodol.*, vol. 105, pp. 1–18, Nov. 2017.
[68] J. Gao, Y. Shen, J. Liu, M. Ito, and N. Shiratori, “Adaptive Traffic Signal Control: Deep Reinforcement Learning Algorithm with Experience Replay and Target Network,” *ArXiv170502755 Cs*, May 2017.
[69] E. National Academies of Sciences, *Signal Timing Manual - Second Edition*, Second. Washington, DC: The National Academies Press, 2015.
[70] S. F. Cheng, M. A. Epelman, and R. L. Smith, “CoSIGN: A Parallel Algorithm for Coordinated Traffic Signal Control,” *IEEE Trans. Intell. Transp. Syst.*, vol. 7, no. 4, pp. 551–564, Dec. 2006.
[71] S. T. G. Fleuren, “Optimizing pre-timed control at isolated intersections,” Doctor of Philosophy, Technische Universiteit Eindhoven, Eindhoven, 2017.
[72] N. Funabikiy, Y. Takefuji, and K. C. Lee, “Comparisons of seven neural network models on traffic control problems in multistage interconnection networks,” *IEEE Trans. Comput.*, vol. 42, no. 4, pp. 497–501, Apr. 1993.
[73] M. B. Trabia, M. S. Kaseko, and M. Ande, “A two-stage fuzzy logic controller for traffic signals,” *Transp. Res. Part C Emerg. Technol.*, vol. 7, no. 6, pp. 353–367, Dec. 1999.
[74] J.-H. Lee and H. Lee-Kwang, “Distributed and cooperative fuzzy controllers for traffic intersections group,” *IEEE Trans. Syst. Man Cybern. Part C Appl. Rev.*, vol. 29, no. 2, pp. 263–271, May 1999.
[75] E. D. Ferreira and P. K. Khosla, “Multi agent collaboration using distributed value functions,” in *Proceedings of the IEEE Intelligent Vehicles Symposium, 2000. IV 2000*, 2000, pp. 404–409.
[76] E. Bingham, “Reinforcement learning in neurofuzzy traffic signal control,” *Eur. J. Oper. Res.*, vol. 131, no. 2, pp. 232–241, Jun. 2001.
[77] M. Wiering, J. Vreeken, J. van Veenen, and A. Koopman, “Simulation and optimization of traffic in a city,” in *2004 IEEE Intelligent Vehicles Symposium*, 2004, pp. 453–458.
[78] A. L. C. Bazzan, “A Distributed Approach for Coordination of Traffic Signal Agents,” *Auton. Agents Multi-Agent Syst.*, vol. 10, no. 1, pp. 131–164, Jan. 2005.
[79] T. H. Heung, T. K. Ho, and Y. F. Fung, “Coordinated road-junction traffic control by dynamic programming,” *IEEE Trans. Intell. Transp. Syst.*, vol. 6, no. 3, pp. 341–350, Sep. 2005.
[80] D. Srinivasan, M. C. Choy, and R. L. Cheu, “Neural Networks for Real-Time Traffic Signal Control,” *IEEE Trans. Intell. Transp. Syst.*, vol. 7, no. 3, pp. 261–272, Sep. 2006.
[81] G. Shen and X. Kong, “Study on Road Network Traffic Coordination Control Technique With Bus Priority,” *IEEE Trans. Syst. Man Cybern. Part C Appl. Rev.*, vol. 39, no. 3, pp. 343–351, May 2009.
[82] C. Cai, C. K. Wong, and B. G. Heydecker, “Adaptive traffic signal control using approximate dynamic programming,” *Transp. Res. Part C Emerg. Technol.*, vol. 17, no. 5, pp. 456–474, Oct. 2009.
[83] A. L. C. Bazzan, D. de Oliveira, and B. C. da Silva, “Learning in groups of traffic signals,” *Eng. Appl. Artif. Intell.*, vol. 23, no. 4, pp. 560–568, Jun. 2010.
[84] I. Alvarez and A. Poznyak, “Game theory applied to urban traffic control problem,” in *2010 International Conference on Control Automation and Systems (ICCAS)*, 2010, pp. 2164–2169.
[85] J. Qiao, N. Yang, and J. Gao, “Two-Stage Fuzzy Logic Controller for Signalized Intersection,” *IEEE Trans. Syst. Man Cybern. Part Syst. Hum.*, vol. 41, no. 1, pp. 178–184, Jan. 2011.

[86] P. G. Balaji and D. Srinivasan, “Type-2 fuzzy logic based urban traffic management,” *Eng. Appl. Artif. Intell.*, vol. 24, no. 1, pp. 12–22, Feb. 2011.
[87] W. Lu, Y. Zhang, and Y. Xie, “A multi-agent adaptive traffic signal control system using swarm intelligence and neuro-fuzzy reinforcement learning,” in *2011 IEEE Forum on Integrated and Sustainable Transportation System (FISTS)*, 2011, pp. 233–238.
[88] S. El-Tantawy, B. Abdulhai, and H. Abdelgawad, “Multiagent Reinforcement Learning for Integrated Network of Adaptive Traffic Signal Controllers (MARLIN-ATSC): Methodology and Large-Scale Application on Downtown Toronto,” *IEEE Trans. Intell. Transp. Syst.*, vol. 14, no. 3, pp. 1140–1150, Sep. 2013.
[89] H. R. Varia, P. J. Gundaliya, and S. L. Dhingra, “Application of genetic algorithms for joint optimization of signal setting parameters and dynamic traffic assignment for the real network data,” *Res. Transp. Econ.*, vol. 38, no. 1, pp. 35–44, Feb. 2013.
[90] Y. Bi, D. Srinivasan, X. Lu, Z. Sun, and W. Zeng, “Type-2 fuzzy multi-intersection traffic signal control with differential evolution optimization,” *Expert Syst. Appl.*, vol. 41, no. 16, pp. 7338–7349, Nov. 2014.
[91] J. Castán, S. Ibarra, and J. Laria, “Sophisticated Traffic Lights Control using Neural Networks,” *Lat. Am. Trans. IEEE Rev. IEEE Am. Lat.*, vol. 13, no. 1, pp. 96–101, Jan. 2015.
[92] K. Jerry, K. Yujun, O. Kwasi, Z. Enzhan, and T. Parfait, “NetLogo implementation of an ant colony optimisation solution to the traffic problem,” *IET Intell. Transp. Syst.*, vol. 9, no. 9, pp. 862–869, 2015.
[93] S. Araghi, A. Khosravi, and D. Creighton, “Intelligent cuckoo search optimized traffic signal controllers for multi-intersection network,” *Expert Syst. Appl.*, vol. 42, no. 9, pp. 4422–4431, Jun. 2015.
[94] Q. Long, J.-F. Zhang, and Z.-M. Zhou, “Multi-objective traffic signal control model for traffic management,” *Transp. Lett.*, vol. 7, no. 4, pp. 196–200, Apr. 2015.
[95] J. B. Clempner and A. S. Poznyak, “Modeling the multi-traffic signal-control synchronization: A Markov chains game theory approach,” *Eng. Appl. Artif. Intell.*, vol. 43, pp. 147–156, Aug. 2015.
[96] R. Kutadinata, W. Moase, C. Manzie, L. Zhang, and T. Garoni, “Enhancing the performance of existing urban traffic light control through extremum-seeking,” *Transp. Res. Part C Emerg. Technol.*, vol. 62, pp. 1–20, Jan. 2016.
[97] M. Papageorgiou, C. Diakaki, V. Dinopoulou, A. Kotsialos, and Y. Wang, “Review of road traffic control strategies,” *Proc. IEEE*, vol. 91, no. 12, pp. 2043–2067, Dec. 2003.
[98] X.-Y. Lu, P. Varaiya, R. Horowitz, D. Su, and S. Shladover, “Novel Freeway Traffic Control with Variable Speed Limit and Coordinated Ramp Metering,” *Transp. Res. Rec. J. Transp. Res. Board*, vol. 2229, pp. 55–65, Sep. 2011.
[99] A. Hegyi, B. De Schutter, and H. Hellendoorn, “Model predictive control for optimal coordination of ramp metering and variable speed limits,” *Transp. Res. Part C Emerg. Technol.*, vol. 13, no. 3, pp. 185–209, Jun. 2005.
[100] Y. Tsobuta, M. Iwata, and H. Kawashima, “Evaluation Study of Inflow Traffic Control by Ramp Metering on Tokyo Metropolitan Expressway - Semantic Scholar,” 2005.
[101] F. Middleham, “The network effects of ramp metering,” in *8th World Congress on Intelligent Transport SystemsITS America, ITS Australia, ERTICO (Intelligent Transport Systems and Services-Europe)*, 2001.
[102] D. Pisarski, “Collaborative ramp metering control : application to Grenoble south ring,” phdthesis, Université de Grenoble, 2014.
[103] K. Bogenberger and A. D. May, “Advanced coordinated traffic responsive ramp metering strategies,” *Calif. Partn. Adv. Transit Highw. PATH*, pp. 7–8, 1999.

[104] I. Steven and J. Chien, "Evaluating Ramp Metering Control Systems Using Microsimulation for US Interstate Highway I-80," 淡江理工學刊, vol. 4, no. 4, pp. 277–292, 2001.
[105] R. Gordon, "Algorithm for Controlling Spillback from Ramp Meters," *Transp. Res. Rec. J. Transp. Res. Board*, vol. 1554, pp. 162–171, Jan. 1996.
[106] M. Papageorgiou, H. Hadj-Salem, and J.-M. Blosseville, "ALINEA: A LOCAL FEEDBACK CONTROL LAW FOR ON-RAMP METERING," *Transp. Res. Rec.*, no. 1320, 1991.
[107] L. Zhang and D. Levinson, "Ramp metering and freeway bottleneck capacity," *Transp. Res. Part Policy Pract.*, vol. 44, no. 4, pp. 218–235, May 2010.
[108] N. Geroliminis, A. Srivastava, and P. Michalopoulos, "A Dynamic-Zone-Based Coordinated Ramp-Metering Algorithm With Queue Constraints for Minnesota's Freeways," *IEEE Trans. Intell. Transp. Syst.*, vol. 12, no. 4, pp. 1576–1586, Dec. 2011.
[109] J.-C. S. Liu *et al.*, *An Advanced real-time ramp metering system (ARMS): the system concept*, vol. 1232. Texas Transportation Institute, 1994.
[110] I. Papamichail, M. Papageorgiou, V. Vong, and J. Gaffney, "Heuristic Ramp-Metering Coordination Strategy Implemented at Monash Freeway, Australia," *Transp. Res. Rec. J. Transp. Res. Board*, vol. 2178, pp. 10–20, Dec. 2010.
[111] O. Rosas-Jaimes and L. Alvarez-Icaza, "Vehicle density and velocity estimation on highways for on-ramp metering control," *Nonlinear Dyn.*, vol. 49, no. 4, pp. 555–566, Sep. 2007.
[112] X. Sun and R. Horowitz, "A localized switching ramp-metering controller with a queue length regulator for congested freeways," in *Proceedings of the 2005, American Control Conference, 2005.*, 2005, pp. 2141–2146 vol. 3.
[113] C. Taylor, D. Meldrum, and L. Jacobson, "Fuzzy Ramp Metering: Design Overview and Simulation Results," *Transp. Res. Rec. J. Transp. Res. Board*, vol. 1634, pp. 10–18, Jan. 1998.
[114] K. Bogenberger and H. Keller, "An evolutionary fuzzy system for coordinated and traffic responsive ramp metering," in *Proceedings of the 34th Annual Hawaii International Conference on System Sciences*, 2001, pp. 10 pp.-.
[115] H. M. Zhang and S. G. Ritchie, "Freeway ramp metering using artificial neural networks," *Transp. Res. Part C Emerg. Technol.*, vol. 5, no. 5, pp. 273–286, Oct. 1997.
[116] C.-H. Wei, "Analysis of artificial neural network models for freeway ramp metering control," *Artif. Intell. Eng.*, vol. 15, no. 3, pp. 241–252, Jul. 2001.
[117] K. Rezaee, B. Abdulhai, and H. Abdelgawad, "Application of reinforcement learning with continuous state space to ramp metering in real-world conditions," in *2012 15th International IEEE Conference on Intelligent Transportation Systems*, 2012, pp. 1590–1595.
[118] A. Fares and W. Gomaa, "Multi-Agent Reinforcement Learning Control for Ramp Metering," in *Progress in Systems Engineering*, Springer, Cham, 2015, pp. 167–173.
[119] A. H. Ghods, A. R. Kian, and M. Tabibi, "A genetic-fuzzy control application to ramp metering and variable speed limit control," in *2007 IEEE International Conference on Systems, Man and Cybernetics*, 2007, pp. 1723–1728.
[120] A. H. Ghods, A. R. Kian, and M. Tabibi, "Adaptive freeway ramp metering and variable speed limit control: a genetic-fuzzy approach," *IEEE Intell. Transp. Syst. Mag.*, vol. 1, no. 1, pp. 27–36, Spring 2009.
[121] D. Nikovski, N. Nishiuma, Y. Goto, and H. Kumazawa, "Univariate short-term prediction of road travel times," 2005, pp. 1074–1079.
[122] Y. Kamarianakis and P. Prastacos, "Forecasting Traffic Flow Conditions in an Urban Network: Comparison of Multivariate and Univariate Approaches," *Transp. Res. Rec. J. Transp. Res. Board*, vol. 1857, pp. 74–84, Jan. 2003.
[123] R. Eglese, W. Maden, and A. Slater, "A Road Timetable to aid vehicle routing and scheduling," *Comput. Oper. Res.*, vol. 33, no. 12, pp. 3508–3519, Dec. 2006.

[124] A. G. Hobeika and Chang Kyun Kim, "Traffic-flow-prediction systems based on upstream traffic," 1994, pp. 345–350.
[125] D. Wild, "Short-term forecasting based on a transformation and classification of traffic volume time series," *Int. J. Forecast.*, vol. 13, no. 1, pp. 63–72, Mar. 1997.
[126] W. Weijermars and E. van Berkum, "Analyzing highway flow patterns using cluster analysis," 2005, pp. 831–836.
[127] G. E. P. Box and G. M. Jenkins, "Some Comments on a Paper by Chatfield and Prothero and on A Review by Kendall," *J. R. Stat. Soc. Ser. Gen.*, vol. 136, no. 3, pp. 337–352, 1973.
[128] M. S. Ahmed and A. R. Cook, "ANALYSIS OF FREEWAY TRAFFIC TIME-SERIES DATA BY USING BOX-JENKINS TECHNIQUES," *Transp. Res. Rec.*, no. 722, 1979.
[129] "AUTOREGRESSIVE INTEGRATED MOVING AVERAGE MODELS (ARIMA)." [Online]. Available: http://www.forecastingsolutions.com/arima.html. [Accessed: 08-May-2018].
[130] H. Chen, S. Grant-Muller, L. Mussone, and F. Montgomery, "A Study of Hybrid Neural Network Approaches and the Effects of Missing Data on Traffic Forecasting," *Neural Comput. Appl.*, vol. 10, no. 3, pp. 277–286, 2001.
[131] F. Hai-Liang, C. Di, L. Qing-Jia, and C. Chun-Xiao, "Multi-Scale Network Traffic Prediction Using a Two-Stage Neural Network Combined Model," in *2006 International Conference on Wireless Communications, Networking and Mobile Computing*, 2006, pp. 1–5.
[132] E. S. Yu and C. Y. R. Chen, "Traffic prediction using neural networks," in *, IEEE Global Telecommunications Conference, 1993, including a Communications Theory Mini-Conference. Technical Program Conference Record, IEEE in Houston. GLOBECOM '93*, 1993, pp. 991–995 vol.2.
[133] J.-M. Xiao and X.-H. Wang, "Study on traffic flow prediction using RBF neural network," in *Proceedings of 2004 International Conference on Machine Learning and Cybernetics, 2004*, 2004, vol. 5, pp. 2672–2675 vol.5.